\documentclass[traditabstract, longauth]{aa} 
\usepackage{txfonts}
\usepackage{graphicx}
\usepackage{caption}
\usepackage{subcaption}
\usepackage{color} 
\usepackage{natbib}
\bibpunct{(}{)}{;}{a}{}{,}
\def\gapprox{{_>\atop{^\sim}}}
\def\lapprox{{_<\atop{^\sim}}}

\def\cmmd{\rm {cm^{-3}}}
\def\cmmt{\rm {cm^{-2}}}

\def\s-1{\rm {s^{-1}}}

\def\HC3N{HC$_3$N}

\def\kms{\hbox{${\rm km\,s}^{-1}$}}
\def\msun{M$_{\odot}$}
\def\lsun{L$_{\odot}$}

\newcommand{\asec}{\mbox{$''$}}

\usepackage{url}

\begin{document}

 \title{ALMA resolves the remarkable molecular jet and rotating wind in the extremely radio-quiet galaxy NGC~1377}

\author{S. Aalto
          \inst{1}
          \and
           N. Falstad\inst{1}
        \and
        S. Muller\inst{1}
          \and
        K. Wada\inst{2}
        \and
          J. S. Gallagher\inst{3}
          \and
         S. K\"onig\inst{1}
          \and
          K. Sakamoto\inst{4}
        \and
        W. Vlemmings\inst{1}
        \and
        C. Ceccobello\inst{1}
        \and
        K. Dasyra\inst{5}
        \and
         F. Combes\inst{6}
         \and
          S. Garc\'ia-Burillo\inst{7}
        \and
        Y. Oya\inst{10}
        \and
         S. Mart\'in\inst{8,9}
        \and    
          P. van der Werf\inst{11}
        \and
        A. S. Evans\inst{12}
        \and
        J. Kotilainen\inst{13}
             }

 \institute{Department of Space, Earth and Environment, Onsala Space Observatory, Chalmers University of Technology, 
              SE-439 92 Onsala, Sweden\\
              \email{saalto@chalmers.se}
\and Kagoshima University, Kagoshima 890-0065, Japan
\and Department of Astronomy, University of Wisconsin-Madison, 5534 Sterling, 475 North Charter Street, Madison WI 53706, USA
\and Institute of Astronomy and Astrophysics, Academia Sinica, PO Box 23-141, 10617 Taipei, Taiwan 
\and Department of Astrophysics, Astronomy \& Mechanics, Faculty of Physics, University of Athens, Panepistimiopolis Zografos 15784, Greece
\and Observatoire de Paris, College de France, PSL University, Sorbonne University, LERMA, CNRS,  Paris
\and Observatorio Astron\'omico Nacional (OAN)-Observatorio de Madrid, Alfonso XII 3, 28014-Madrid, Spain
\and European Southern Observatory, Alonso de Córdova 3107, Vitacura, Santiago 763-035, Chile
\and Joint ALMA Observatory, Alonso de Córdova 3107, Vitacura, Santiago 763-035, Chile
\and Department of Physics, The University of Tokyo 7-3-1, Hongo, Bunkyo-ku, Tokyo 113-0033, Japan
\and Leiden Observatory, Leiden University, 2300 RA, Leiden, The Netherlands
\and University of Virginia, Charlottesville, VA 22904, USA, NRAO, 520 Edgemont Road, Charlottesville, VA 22903, USA
\and Finnish Centre for Astronomy with ESO (FINCA), University of Turku Quantum, Vesilinnantie 5, FI-20014 University of Turku, Finland
 }

   \date{Received xx; accepted xx}

  \abstract{Submillimetre and millimetre line and continuum observations are important in probing the morphology, column density, and dynamics of the molecular gas and dust
around obscured active galactic nuclei (AGNs) and their mechanical feedback. With very high-resolution ($0.\asec 02 \times 0.\asec 03$ ($2 \times 3$ pc)) ALMA 345~GHz
observations of CO 3--2,  HCO$^+$ 4--3, vibrationally excited HCN 4--3 $\nu_2$=1$f$, and continuum we have studied the remarkable, extremely radio-quiet,
molecular jet and wind of the lenticular galaxy NGC~1377. The outflow structure is resolved, revealing a 150 pc long, clumpy, high-velocity ($\sim$600 \kms), collimated molecular jet
where the molecular emission is emerging from the spine of the jet with an average diameter of 3-7 pc. The jet widens to 10-15 pc about 25 pc from the centre,  which is possibly due to jet-wind interactions.
A narrow-angle (50$^{\circ}$-70$^{\circ}$), misaligned and rotating molecular wind surrounds the jet, and both are enveloped by a larger-scale CO-emitting structure at near-systemic velocity. 
The jet and narrow wind have steep radial gas excitation gradients and appear turbulent with high gas dispersion ($\sigma$$>$40 \kms).  The jet shows velocity reversals that we propose are caused by
precession, or more episodic directional changes.  We discuss the mechanisms powering the outflow, and we find that an important process for the molecular jet and narrow wind is likely magneto-centrifugal driving. 
In contrast, the large-scale CO-envelope may be a slow wind, or cocoon that stems from jet-wind interactions.
An asymmetric, nuclear $r$$\sim$2 pc dust structure with a high inferred molecular column density $N$(H$_2) \simeq 1.8 \times 10^{24}$$\cmmt$ is detected in continuum and also shows compact
emission from vibrationally excited HCN. The nuclear dust emission is hot ($T_{\rm d}$$>$180 K) and its luminosity is likely powered by a buried AGN. The lopsided structure appears to be a warped disk, 
which is responsible for a significant part of the nuclear obscuration and possibly formed as a result of uneven gas inflows. The dynamical mass inside $r$=1.4 pc is estimated to $9^{+2}_{-3} \times 10^6$ \msun, implying
that the supermassive black hole (SMBH) has a high mass with respect to the stellar velocity dispersion of NGC~1377.
We suggest that the SMBH of NGC~1377 is currently in a state of moderate growth, at the end of a more intense phase of accretion and also evolving from a state of more extreme nuclear obscuration. 
The nuclear growth may be fuelled by low-angular momentum gas inflowing from the gas ejected in the molecular jet and wind. Such a feedback-loop of cyclic outflows and central accretion could explain
why there is still a significant reservoir of molecular gas in this ageing, lenticular galaxy. A feedback-loop would be an effective process in growing the nuclear SMBH and thus would constitute an important phase
in the evolution of NGC 1377. This also invites new questions as to SMBH growth processes in obscured, dusty galaxies.
}


 \maketitle

\section{Introduction}
\label{s:intro}

Feedback in the form of outflows and winds is an important process in regulating both the evolution of the stellar constituent
of galaxies and the growth of their nuclear, supermassive black holes (SMBHs). Studying the link between accretion and the feedback of SMBHs (=active galactic nuclei, AGN) 
as well as star formation  is essential for our understanding of how black holes and host galaxies evolve together  \citep[e.g.][]{ho04}.

The most active growth phase is suspected to commence when the SMBH is deeply dust enshrouded \citep[e.g.][]{lusso13, kocevski15}, but
please see \citet{ricci17} for an additional discussion on the link between obscuration and accretion. 
Some of the nuclear obscuration is believed to occur in the form of a parsec-scale dusty torus which the Atacama Large Millimeter/submm Array (ALMA) is now able to image
\citep[e.g.][]{garcia16,gallimore16,aalto17,combes19,impellizzeri19,garcia19}.  Feedback from the AGN eventually clears its surroundings and AGN outflows can eject
large amounts of molecular gas from galaxy nuclei \citep[e.g.][]{cicone14,fiore17,veilleux20}.  In some cases the feedback impacts the evolution of the galaxy 
on time scales of only a few tens of Myr. The fate of the gas in the outflow is an important factor in determining whether the feedback turns off
the growth of the galaxy for a long time, or whether the evolution is simply suspended. One question is if the gas in the outflow will leave the galaxy, or if it will return to support
further growth \citep[e.g.][]{aalto12b,pereira18, fluetsch19}.

The AGN feedback is presumed to turn off the nuclear growth, for example, by removing gas in the vicinity of the activity,
but also through radiative feedback that heat the nuclear region preventing further growth. Recently there have been
ideas that an outflow may also provide positive feedback in the form of star formation \citep{maiolino17}. Jets may also aid accretion
through removing angular momentum from the gas in the inner region, allowing the remaining gas to reach the SMBH \citep[e.g.][]{blandford82,aalto16}. Cold gas
is an important fuel for SMBH accretion, since hot gas feeding is difficult, and it may occur via inflows along the disk in a bar or spiral arms, or
through gas returning in a fountain,  that is from gas decelerating in an outflow, stripped of its angular momentum. The importance of positive feedback for galaxy evolution
is not well studied and requires further observational attention.

NGC~1377 is a nearby (21 Mpc (1\arcsec=102~pc)) lenticular galaxy with a far-infrared (FIR) luminosity of $L_{\rm FIR}$=$4.3 \times 10^9$ L$_{\sun}$ \citep{roussel03}.
It is extremely radio-quiet and significantly off  the radio-FIR correlation \citep{helou85}; its radio emission is weaker (with respect to its FIR emission) by a
factor $\approx$37 compared to normal galaxies \citep{roussel03,roussel06, costagliola16}.  The central region is dust-enshrouded \citep[e.g.][]{spoon07} and the source of the FIR luminosity (and the cause of
 its radio deficiency) has remained elusive. A nascent starburst \citep{roussel03,roussel06} or a radio-quiet AGN \citep{imanishi06,imanishi09} have been proposed as
 possible solutions.  \citet{costagliola16} detect faint radio emission with a synchrotron spectrum and  estimate a star formation rate (SFR) $<$0.1 \msun\ yr$^{-1}$ based on the radio emission.
This is not sufficient to power the observed IR luminosity and supports the notion of a radio-quiet AGN.
A powerful molecular outflow with a mass outflow rate of  8 -- 35 \msun\ yr$^{-1}$ was found with the Submillimeter Array (SMA) \citep{aalto12b}. ALMA observations revealed that the high-velocity gas is
in the form of a collimated outflow - a molecular jet with a high momentum flux of $>$14 $L/c$. Velocity reversals along the molecular jet are suggested to be indications of precession \citep{aalto16}. 
ALMA also revealed a slow wind along the minor axis and the relation between the molecular jet and wind is not clear.
 
Although the high degree of collimation, combined with the large momentum flux, point towards an AGN,  the nature of the power source behind the molecular outflow, and the FIR luminosity,  is still
not fully understood.  Jets are indeed generally identified with accretion \citep{blandford98} consistent with an AGN. However, it is not clear if the
molecular gas is carried out by an extremely faint radio-jet powered by inefficient, hot ("radio-mode") accretion, or if it is driven by effective accretion of cold gas ("quasar-mode"). In the latter case,
the molecular jet would represent an unexplored form of quasar-mode feedback, where the high-velocity gas is expelled in a collimated outflow and not a wide-angle wind \citep{veilleux13}. It is also
possible that the outflow may be powered by accretion onto the nuclear disk, and not onto the SMBH. This would then be a new form of outflow not previously seen in galactic-scale outflows.

Observations at mid-infrared (mid-IR) wavelengths reveal a compact ($<$0.\asec 14), high surface brightness source \citep{imanishi11} in the nucleus of NGC~1377 suggesting hot dust. This is
consistent with the presence of nuclear vibrationally excited HCN emission which requires a 14\,$\mu$m background temperature of $T_{\rm B}(14\, \mu{\rm m})$$>$100 K to be excited \citep{aalto15b,aalto16}. 
It is not clear whether this structure is part of an obscuring torus or disk, and whether it is opaque enough to absorb X-rays emerging from an accreting SMBH. It is also not understood whether (or how)  the obscuring material
is linked to the outflowing gas in the jet. High-resolution CO 6--5 observations \citep{aalto17} reveal warm ($T$$>$100 K) nuclear gas at the base of the molecular outflow and also an apparent circumnuclear disk (CND) or
torus of radius $r$=2.5 pc, but no nuclear submm continuum is detected. 

In this paper we present new high resolution (0.\asec 025 $\times$ 0.\asec 033) (2.5 $\times$ 3 pc) ALMA CO $J$= 3--2, HCO$^{+}$ $J$=4--3, HCN $J$=4--3 $\nu_2$=1$f$ and also H$^{13}$CN $J$=4--3 of NGC~1377. The paper is
organised as follows: In Sec.~\ref{s:obs} we present the observations and in Sec.~\ref{s:results} we show the results in the form of moment maps as well as images of the high-velocity, and low-velocity, CO 3--2 emission. 
We also present the 0.8~mm continuum image.
In Sec.~\ref{s:physical} we present the physical conditions (temperature and dispersion) of the gas and dust in nucleus and outflow components.
In Sec.~\ref{s:nucleus} we discuss the nuclear properties of NGC~1377 including the gas and dust column density, nuclear dynamics and enclosed mass, evidence for a gas inflow and a discussion of the luminosity source.
We discuss the energetics and turbulence of the jet and wind in Sec.~\ref{s:energetics} and  in Sec.~\ref{s:origin} we discuss the origin and possible driving mechanisms of the jet and wind. In Sec.~\ref{s:disk} we briefly discuss
the larger scale disk rotation and mass. Finally we place the properties of the jet and wind of NGC~1377 in the context of its nuclear growth (Sec.~\ref{s:growth}).


\section{Observations}

\label{s:obs}

Observations of the CO $J$=3--2 line (at $0.\asec 25 \times 0.\asec 18$ resolution) have been previously obtained with the Atacama Large Millimeter/submm Array (ALMA)
in August 2014 and reported in \citet{aalto16}. Here we present new ALMA observations, with the same spectral tuning but with more extended array configurations, taken in
November 2017.   A journal of the observations, including the 2014 lower resolution observations,  is given in Table~\ref{tab:obs}. The phase centre was set to
R.A.=03h36m39.074, Dec.=$-$20d54m07.0553 (ICRS).

The correlator was set up with four 1.875~GHz-wide spectral windows centred at 342.2, 343.8 (covering the CO $J$=3--2 line, rest frequency
345.79599~GHz), 354.7 (covering the HCO$^+$ $J$=4--3 and HCN $J$=4--3 $\nu$=1$f$,  lines, rest frequencies 356.73422 and 356.25557~GHz, respectively), and 356.1~GHz, 
with a native channel spacing of $\sim$2~MHz. 

After calibration within the CASA reduction package, the visibility set was imported into the AIPS
package for further imaging. The synthesised beam is $0.\asec 030 \, \times \, 0.\asec 022$ (3$\times$2.2 pc for NGC~1377) with position angle PA=59$^{\circ}$.
With Briggs weighting (parameter robust set to 0.5),  the resulting data has a sensitivity of 0.36~mJy per beam in a 10~\kms\ (12~MHz) channel width.

\begin{table*}[h]
\caption{Journal of the observations.}
\label{tab:obs}
\begin{center} \begin{tabular}{cccccccc}
    \hline
Date of      & N$_{\rm ant}$  & PWV $^{(c)}$ & t$_{\rm on}$ $^{(d)}$ & B$_{\rm min}$ / B$_{\rm max}$ $^{(e)}$ & Bandpass        & Flux       & Gain \\
observations & $^{(b)}$      & (mm) & (min)          & (m / km)       & calibrator     & calibrator ($^{(f)}$ Jy) & calibrator \\
    \hline

12 Aug 2014 $^{(a1)}$ & 33 & $\sim$0.5  & 30             & 31        / 1.1        & J\,0423$-$0120 & J\,0334$-$401 ($0.46 \pm 0.14$) & J\,0340$-$2119 \\ 
12 Nov 2017 $^{(a2)}$ & 45 & $\sim$0.6 & 50 & 113 / 13.9 & J\,0522$-$3627 & J\,0522$-$3627 ($4.81 \pm 0.23$) & J\,0340$-$2119 \\ 
21 Nov 2017 $^{(a2)}$ & 47 & 0.4--0.6 & 50 & 92 / 8.5 & J\,0522$-$3627 & J\,0522$-$3627 ($4.70 \pm 0.19$) & J\,0340$-$2119 \\ 

\hline
\end{tabular} \end{center}
 \mbox{\,} \vskip -.5cm
 $(a1)$ ALMA project number 2012.1.00900.S; $(a2)$ ALMA project number 2017.1.0659.S;
 $(b)$ Number of 12\,m-antennas in the array;
 $(c)$ Amount of precipitable water vapour in the atmosphere;
 $(d)$ On-source time;
 $(e)$ Respectively minimum and maximum projected baseline;
 $(f)$ Flux density at 350~GHz, as retrieved from the ALMA flux monitoring database.
\end{table*}



\section{Results}
\label{s:results}

\subsection{CO emission features: Jet, wind, and disk}
\label{s:features}

The CO emission delineates several apparently separate features: 

\begin{description}
\item  {\bf a.} A well collimated  structure which essentially is visible at high velocities. We interpret the high-velocity feature as a molecular jet (see \citet{aalto16}) and
we refer to it as such throughout this paper.

\item {\bf b.} Minor axis emission at intermediate-to-low (30 to 80 \kms) velocities, a v-shaped structure which surrounds the high-velocity jet - in particular to the north of the nucleus and most prominent
on the redshifted, eastern side. We refer to this structure as the narrow wind.

\item {\bf c.} Wide (width of $\sim$2\asec)), minor-axis emission at low (20-30 \kms projected) velocities. The velocity shift between north and south is very small. We refer to this feature as the slow wind (or
cocoon (Sec.~\ref{s:bow})). 

\item {\bf d. }There is also gas aligned with the stellar thick disk (on scales out to $r$=60-100 pc) and we refer to this as the disk, although, as we shall see, it is unclear if the gas is actually in a rotating disk. 

\item {\bf e.} Gas in the inner $r$=5 pc is denominated the nucleus. 

\item {\bf f.} There are curved structures along the minor axis, generally perpendicular to the outflow. These features may be  bow shocks and we refer to them as such in the paper. 
\end{description}

\begin{figure*}[tbh]
\includegraphics[scale=0.7]{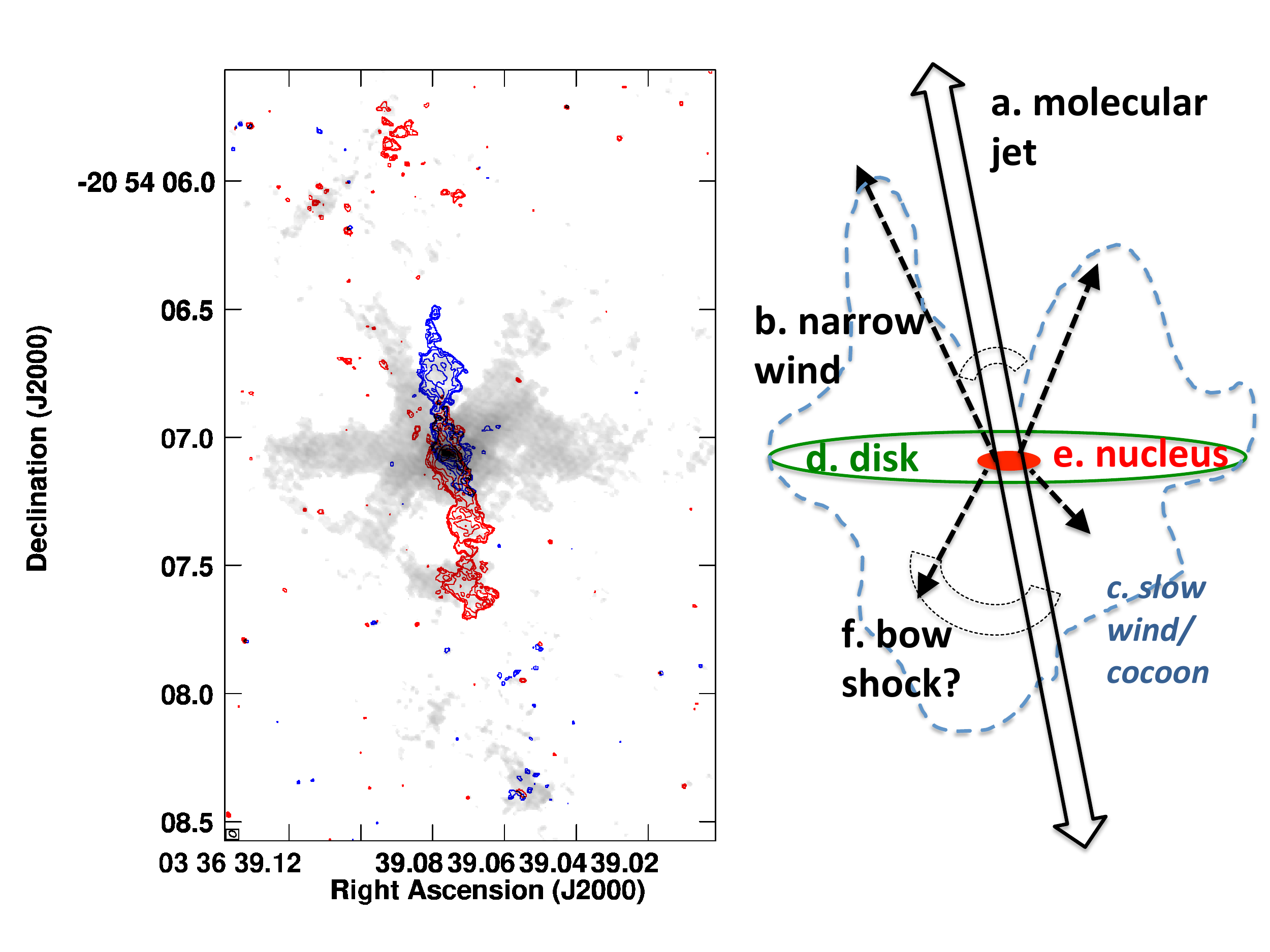}
\caption{\label{f:jet_sys} Left panel: CO 3--2 integrated intensity image where emission close to systemic velocity (0-70 \kms) is shown in greyscale (ranging from
0 to 1 Jy \kms). The high-velocity ($\pm$80 to $\pm$160 \kms) emission from the molecular jet is shown in contours (with the red and blue showing the velocity reversals). 
The contour levels are $6\times 10^{-3}$ $\times$(1,2,4, 8, 16,32,64)  Jy \kms\ beam$^{-1}$.  Right: Chart of the various components of the molecular structure of NGC~1377 as presented
in Sec.~\ref{s:features}. The outline of the slow wind ( which may be a jet cocoon (Sec.~\ref{s:bow})) is largely taken from the low-velocity gas in \citet{aalto16} since a significant fraction of 
he extended low-velocity gas is missing in the high-velocity data (Sec~\ref{s:moment}). }
\end{figure*}

To help the reader navigate the complex structure of the molecular gas of NGC~1377, we present the various regions in a cartoon "finder-chart" in the right panel of Fig.~\ref{f:jet_sys}.
In \citet{aalto12b} the systemic velocity was suggested to be $v_{\rm sys}$=1740 \kms\  based on the shape of the CO 2--1 line. Here we find a $v_{\rm sys}$=1730 \kms\ of the nuclear rotation (optical velocity
definition, see Sec.~\ref{s:hivel}) and we define high-velocity gas as $v > 80$ \kms\ (dominated by the jet structure and the nucleus). We note however, that the jet also contains lower velocity emission and that 
high-velocity gas is also found in the nucleus.

\subsection{CO 3--2 moment maps}
\label{s:moment}
In the high resolution data we recover $\sim$60\% of the emission in the previous, $0.\asec 25 \times 0.\asec 18$ resolution CO 3--2 map \citep{aalto16}. Most of the missing flux originates in
extended, low-velocity emission associated with the slow wind  (feature c in Fig.~\ref{f:jet_sys}). For the compact and collimated structures, such as the jet and narrow wind, we recover
all of the previous flux. In this paper we focus on the small-scale structure of the jet, wind and disk of NGC~1377 and therefore settle with presenting the high resolution results only,
without merging with the older, lower resolution data set.

\begin{figure*}[tbh]
\includegraphics[scale=0.46]{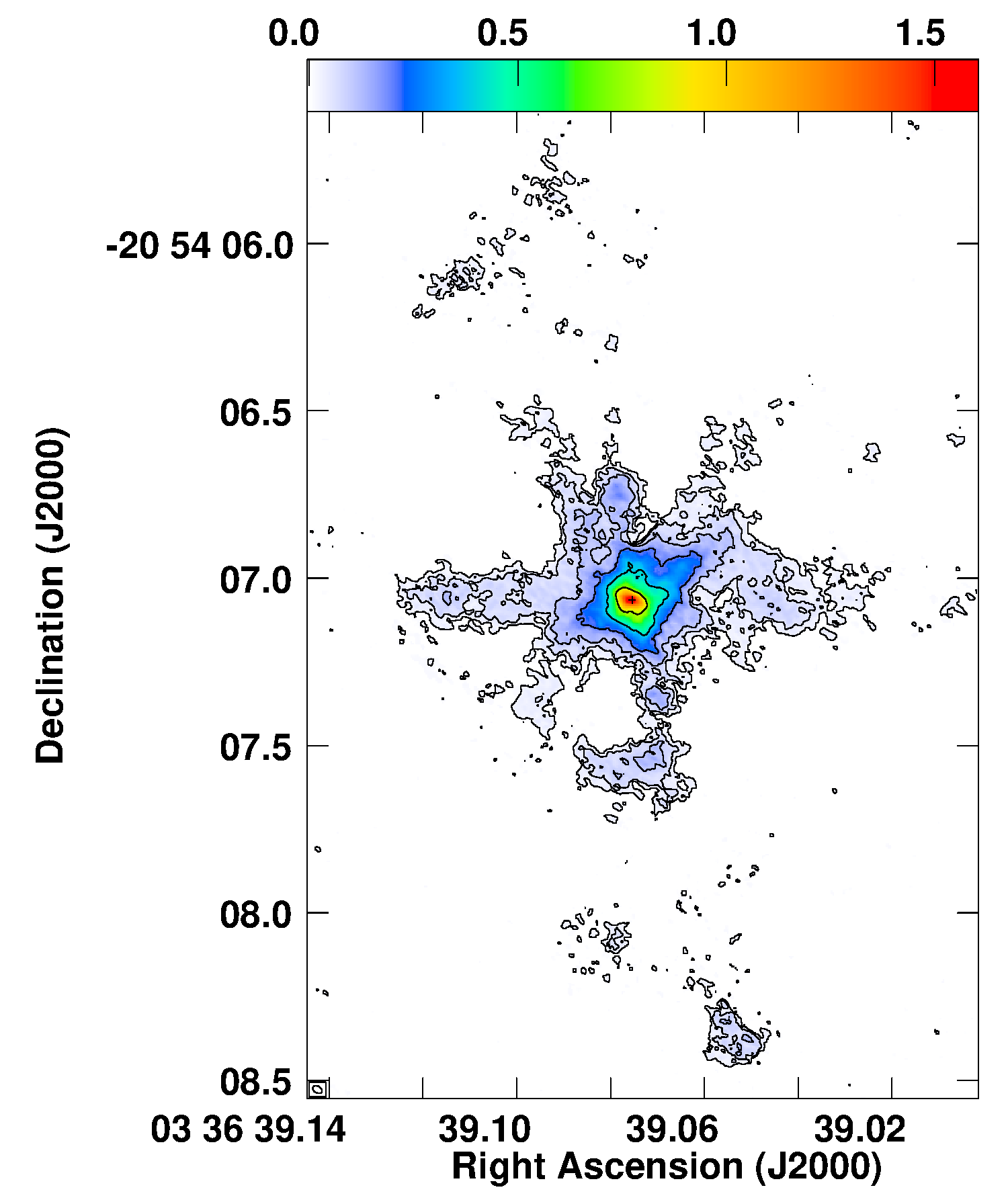}
\includegraphics[scale=0.46]{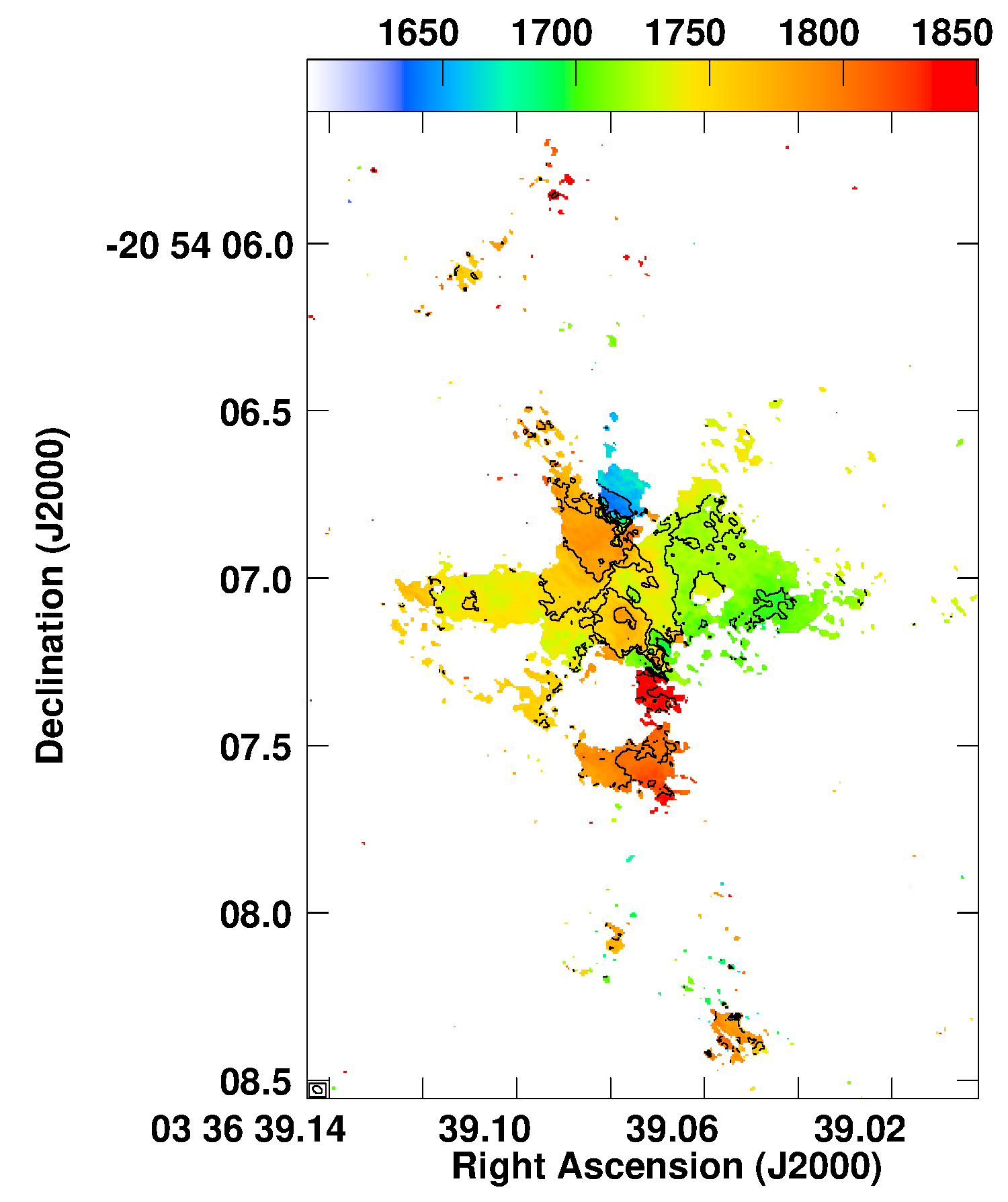}
\caption{\label{f:mom} Left: Integrated CO 3--2 intensity (mom0) with contours 0.017$\times$ (1, 3, 6, 12, 24, 48)  Jy \kms\ beam$^{-1}$. Colours range
from 0 to 1.7 Jy \kms\ beam$^{-1}$. Right: Velocity field (mom1) with contours from 1675 \kms\ to 1825 \kms\ (steps of 25 \kms), colours from 1650 to 1850 \kms. 
The cross indicates the position of the 345 GHz continuum peak.
}
\end{figure*}

\begin{figure}[tbh]
\includegraphics[scale=0.46]{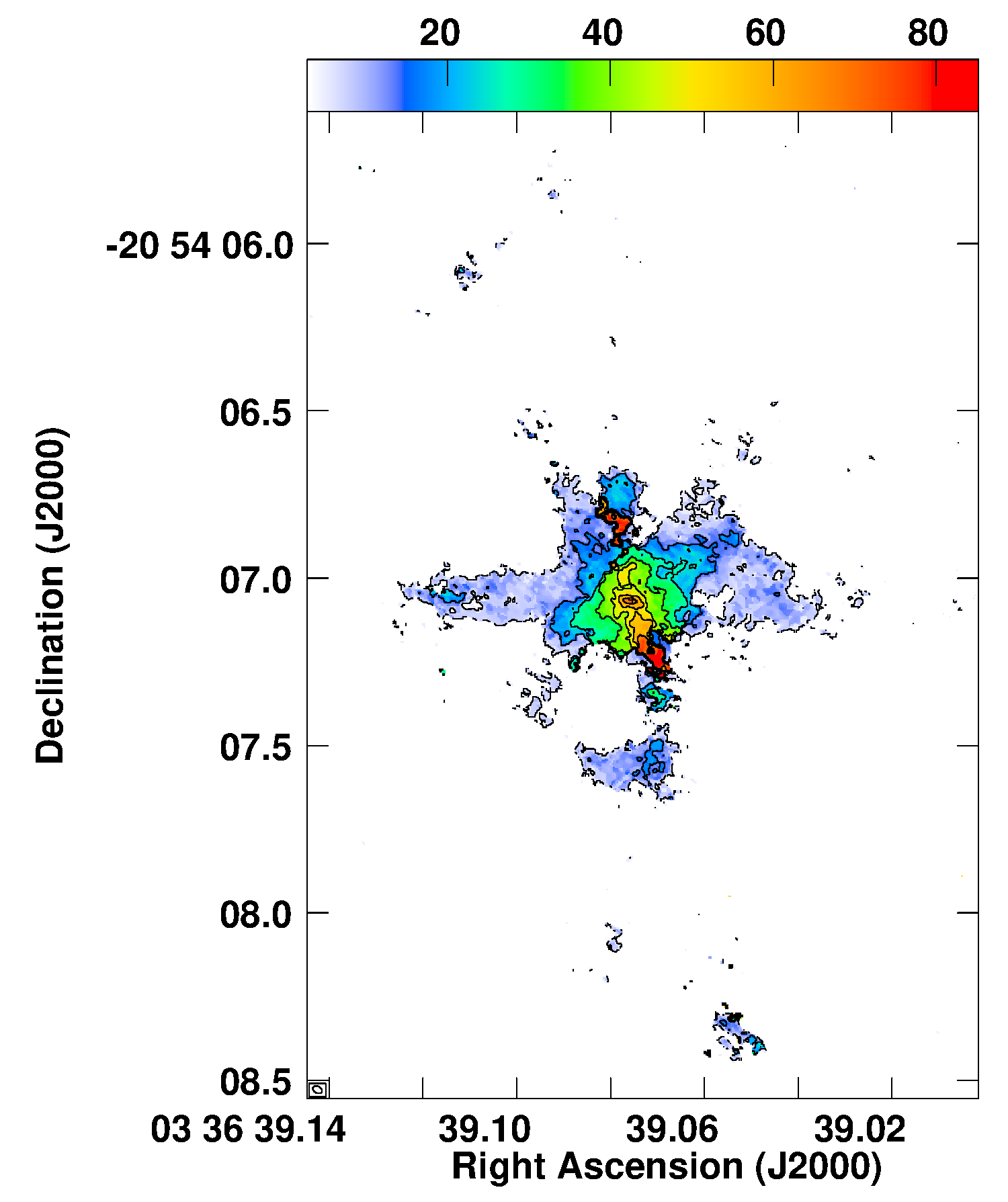}
\caption{\label{f:mom2} Dispersion map (mom2) with contours 
5$\times$(1, 3, 5, 7, 9, 11, 13) \kms. Colours range from 3 to 85 \kms. The cross indicates the position of the 345 GHz continuum peak.
}
\end{figure}

The CO 3--2 integrated intensity (moment 0) map and velocity field (moment 1) are presented in Fig~\ref{f:mom} and the dispersion map (moment 2) in Fig.~\ref{f:mom2}.
We smoothed to two channel resolution, then for the moment 0 map we clipped at the 3$\sigma$ level, and for the moment 1 and
2 maps we clipped at 4$\sigma$. The velocity centroids were determined through a flux-weighted first moment of the spectrum of each pixel, therefore
assigning one velocity to a spectral structure. The dispersion ($\sigma$) was determined through a flux-weighted second moment of the spectrum of each pixel.
This corresponds to the one dimensional velocity dispersion (i.e. the FWHM line width ($\Delta V$ ) of the spectrum divided by 2.35 for a Gaussian line profile).

\subsubsection{Integrated intensity map}
\label{s:mom0}

Bright CO 3--2 emission is found in the inner 0.\asec 3 (30 pc) with an elongation along the minor axis. Intense emission is also located
0.\asec 5 along the molecular jet (with a position angle (PA) of $\sim$11$^{\circ}$ \citep{aalto16}), in the eastern part of a (possible) molecular disk aligned with the PA=90$^{\circ}$ stellar major axis, and
also in filamentary north-western and south-eastern minor axis structures. Extended emission can be found in the molecular jet (with a total size of 3\asec) and in
v-shaped (or an asymmetric cross) minor axis emission of opening angle  $\theta$=50$^{\circ}$-$70^{\circ}$.  Faint emission is also found out to $r$=0.\asec 6 along the stellar major axis. 
The CO 3--2 brightness temperatures and their interpretation in terms of gas temperatures are presented in Sec.~\ref{s:physical}.

\subsubsection{Velocity field}
\label{s:mom1}

Fig~\ref{f:mom} shows that the CO 3--2 velocity field is complex with the highest velocities in the molecular jet which is also showing radial velocity reversals. Lower velocity shifts are seen along the stellar major
axis (25 \kms\ east-west velocity shift (at PA=90$^{\circ}$) observed before by \citet{aalto16}) and  between east-west components in the narrow wind.  Abrupt velocity changes from the red-shifted narrow wind to the
blue-shifted jet  can be seen  0.\asec 25 to the north. This is likely  evidence of dynamically separate structures overlapping along the line-of-sight.  
The curved feature from the southern jet to the east shows a velocity gradient with receding velocities as it connects with the narrow wind, and we also detect a velocity gradient across the northern "expansion" structure
in the jet. The velocity field in the central 0.\asec 1 has a similar PA of 140$^{\circ}$ that was found for the high frequency (690 GHz) band 9 central velocity field \citep{aalto17}.  However, the
intensity weighted moment 1 map is not sensitive to the highest velocity emission, and in Sec.~\ref{s:highest_vel} below we find a different PA of the velocity shift as we get closer to the nucleus.

\subsubsection{Velocity dispersion and line widths}
\label{s:mom2}

High dispersion emission ($\sigma$=40-90 \kms) is found in the nucleus and along the molecular jet out to a distance of $\sim$0.\asec 25. 
Dispersion is generally dropping with radius in all structures. After the "expansion",  the jet $\sigma$ drops significantly, down to values of 8-15 \kms\ in the
north. To the south there is a similar, but less pronounced, trend. Dispersion in the narrow wind is $\sigma$$\sim$40 close to the nucleus, and then drops to  $\sigma$=10-15 \kms\ after 25-30 pc.  
Emission of intermediate $\sigma$=10-30 \kms\  is generally found on size scales 0.\asec 25 to 0.\asec 5, and extending to the east along the stellar major axis and in the eastern part of the minor axis.
A discussion of $\sigma$ in terms of physical conditions and turbulence can be found in  Sec.~\ref{s:physical}.


\subsection{High-velocity CO 3--2 emission}
\label{s:hivel}

\subsubsection{The highest velocity gas}
\label{s:highest_vel}
 
\begin{figure}[tbh]
\includegraphics[scale=0.35]{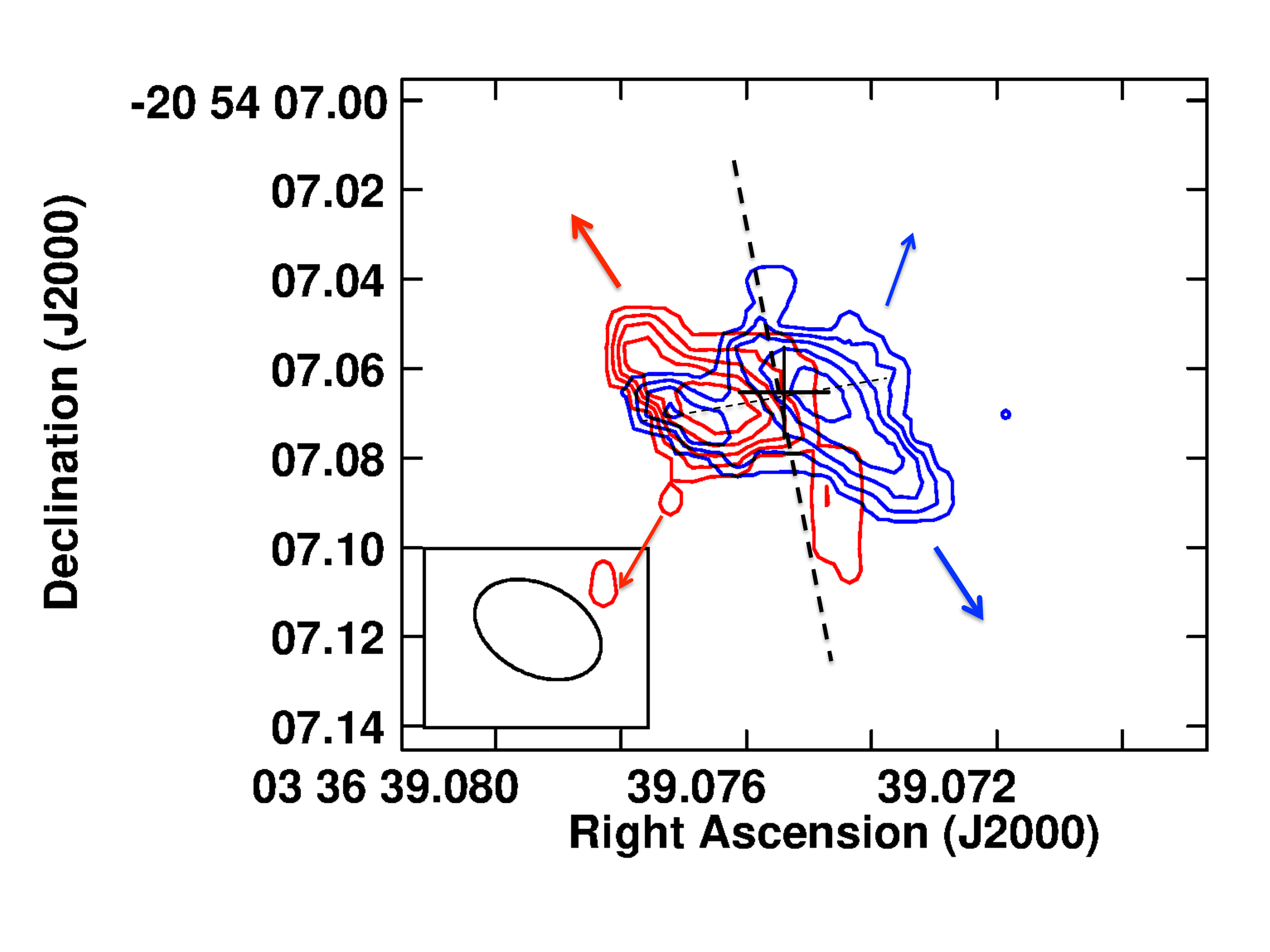}
\caption{\label{f:rot}  Red- and blueshifted high-velocity gas:  Nuclear orientation of rotation (Blue: 1552 - 1572  \kms; Red: 1886 - 1906 \kms). (Contours 0.004$\times$(1,2,3,4,5) 
Jy beam$^{-1}$ kms$^{-1}$). The thick, black, and dashed line indicates the jet orientation, the thin line the orientation of the nuclear velocity shift. The red and blue arrows indicate the
narrow wind.
}
\end{figure}

The highest velocity emission $>$160 \kms, is found in the nucleus. Red- and blue-shifted velocities of 160 \kms\  peak in the nucleus with a 
PA of  $\sim$100$^{\circ}$ and $\Delta$x of 0.\asec 028 $\pm$ 0.\asec 003  (2.8$\pm$0.3 pc)(Fig.~\ref{f:rot}).  The centre of this velocity shift is at $\sim$1730 \kms\ and the  kinematic centre
 is slightly offset to the east from the continuum peak (Fig.~\ref{f:rot}). In Sec.~\ref{s:dyn} we discuss this velocity shift as representing a nuclear rotating disk 
 and estimate its dynamical mass. We also find extensions of the high-velocity gas that can be linked to the narrow wind and jet (Sec.~\ref{s:origin}).
 In Fig.~\ref{f:rot} another, blue-shifted high-velocity component is visible 2 pc east of the continuum peak. The origin of this emission is not clear, but it is difficult to
 assign it to any outflowing structure. It is located where the asymmetric continuum emission (Sec.~\ref{s:continuum}) has an extension to the east. 

\subsubsection{The molecular jet}

High-velocity gas (projected $v$=80-160 \kms) (Fig. ~\ref{f:jet_sys}) is aligned in a $\pm$1.\asec 5 ($\pm$150 pc) long, collimated, jet. It has a symmetry angle of PA=11$^{\circ}$ \citep{aalto16}, and the jet
structure is resolved in our 2 by 3 pc beam. The jet width is 3-7 pc on average, but widens to 10-17  (Fig. ~\ref{f:jet_sys}) about 20 to 25 pc north of the nucleus. The emission remains
wide for $\sim$20 pc before narrowing again, and is also diverging from the symmetry axis of the jet. The widening and narrowing of the jet occurs further out (40 pc) to the south, but the emission
departs from the symmetry axis at roughly the same distance from the centre. There is a gap in high-velocity emission along the jet until after 0.\asec 5 where emission appears again, at the reverse velocities. 
In the north, the high-velocity gas becomes redshifted, and vice versa to the south. This behaviour was observed at lower spatial resolution and discussed by \citet{aalto16}. Also, the emission close to the nucleus
to the north shows wiggles on small scales.

In a precessing jet scenario \citep{aalto16}, there is also jet emission at lower velocity which should be further away from the jet symmetry axis. In Fig.~\ref{f:jet_channel} we show how the structure and width is
changing with velocity, where the highest velocity gas is aligned with the jet axis, while at lower projected velocities, the emission is broader and also starts to deviate from the symmetry axis. At lower jet velocities,
curved structures, roughly perpendicular to the jet axis, also become apparent. In particular to the south.

The jet is likely launched from within a small radius (Sec.~\ref{s:origin}) and potential rotation may not be resolved. In Fig.~\ref{f:jet_channel} panels c) and d) show redshifted collimated emission
at lower jet velocities to the north (and vice-versa to the south). It is possible that this is unresolved jet-rotation, but it may also be emission belonging to the narrow wind (see below). There is a also velocity shift (opposite to disk
rotation) in the northern jet expansion which we attribute to internal working surfaces and bow shocks (see Sec.~\ref{s:bow}).

\subsubsection{The narrow wind}
\label{s:narrow}

\begin{figure}[tbh]
\centering
\includegraphics[width=9cm, trim = 0.9cm 0.3cm 10cm 0.1cm, clip]{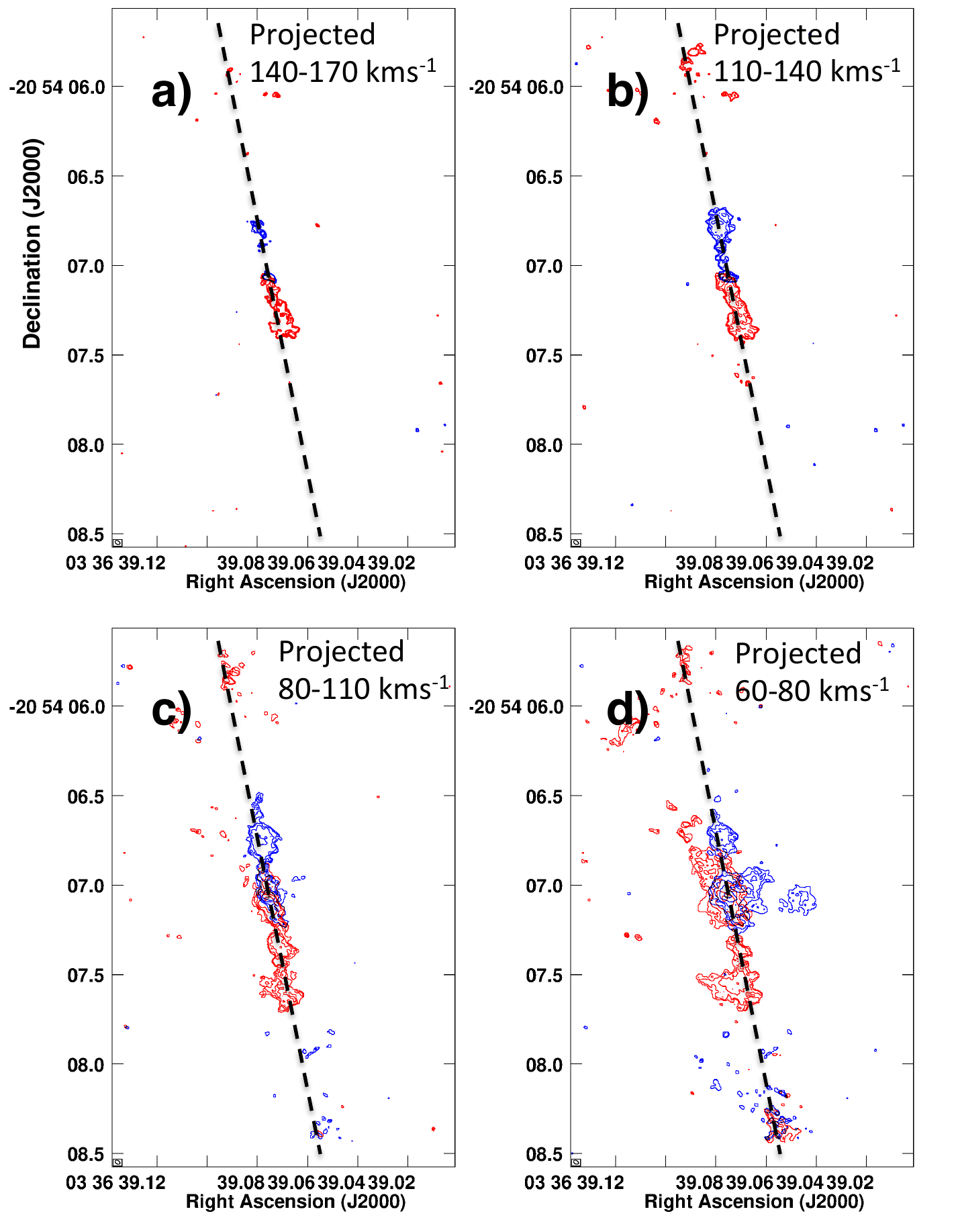}
\caption{\label{f:jet_channel}  Panels showing the structure of the jet at various velocity intervals: Top left panel: Highest velocity, 
projected velocities 140-170 \kms. Top right: High-velocity, projected velocities 110 -140 \kms. Bottom left: Intermediate velocity, projected velocities 80-110 \kms. Bottom right: Lower velocity
60-80 \kms. The black dashed line shows the orientation of the jet axis. Contour levels are (for the blue- and redshifted emission respectively):  Panel a) 0.004$\times$(1,2,3,4,5); Panel b)
0.004$\times$(1,2,4,8,16,32); panels c) and d) 0.004$\times$(1,5,10,20,50) Jy beam$^{-1}$ kms$^{-1}$. }
\end{figure}

In Fig.~\ref{f:jet_sys} the narrow wind is visible as a cross-shaped minor axis emission structure (dominated by the northern emission) of opening angle  $\theta$=50$^{\circ}$-$70^{\circ}$. The wind becomes
apparent in  panels c and d in Fig.~\ref{f:jet_channel} for velocities 60-110 \kms, (but is also apparent at lower velocities) and is also obvious in the moment maps (Fig~\ref{f:mom}). Higher velocity gas in the wind is also seen on small scales
in the nucleus (Fig.~\ref{f:rot}). The wind has a north-south asymmetry in extent of its high-velocity component which extends only 10-20 pc to the south while low-velocity wind emission
can be found on the south-eastern side down to 50 pc.   On the south-western side even the low-velocity wind component is missing. To the north, the eastern component of the wind extends out to 100 pc while the western
part appears to curve down towards the disk at a distance 25 pc from the centre (lower velocity gas extends out to 50 pc). The narrow wind is also not symmetric about the jet axis  (eastern red component is closer to the jet than the blue). 
On the south side of the jet, 0.\asec 5 from the centre, the jet connects to a lane of emission extending to the east and curving north. 

A velocity shift between the east- and west part of the narrow wind is evident in Fig.~\ref{f:rot} and in Fig.~\ref{f:jet_channel} (panels c and d), and appears to be consistent with rotation. 
Velocity shifts between the eastern (red) and western (blue) part of the wind appear to peak at $\pm$110 \kms\ out in the wind, but are as high as $\pm$160 \kms\ near the nucleus.  In \citet{aalto17} we find that the 
690 GHz CO 6--5 emission close to the nucleus also show the red- and blue-shifted v-shaped emission at the base of the narrow wind. In Sec.~\ref{s:origin} we present a
simple model of the wind rotation as a possible signature of a magneto-centrifugal disk-wind.

\subsection{Systemic and low-velocity gas}
\label{s:lovel}

The systemic and low-velocity gas (projected velocities 0-50 \kms)  (Fig~\ref{f:jet_sys}) consists of a bright central feature and  larger scale emission
extending primarily along the minor axis of NGC~1377.  Some low-velocity emission is associated with the narrow wind, but most is wider and extends out to distances
of 75-120 pc. The missing flux in the high resolution map (compared to the $0.\asec 25 \times 0.\asec 18$ resolution CO 3--2 map \citep{aalto16})
is largely in the low-velocity gas.

 \citet{aalto16} reported that the low-velocity emission surrounds the molecular high-velocity jet in a butterfly-like pattern. In this high resolution
map the low-velocity emission that we recover is located in narrow filamentary-like structures. The low-velocity minor axis flow shows a small east-west velocity gradient
(10-30 \kms) and there is also a modest north-south velocity shift of $\sim$20-30 \kms. Low-velocity emission is also associated with the PA=90$^{\circ}$ stellar disk of NGC~1377
out to a radius of $r$=60-100 pc.  As noted in Sec.~\ref{s:mom1}, the velocity shift along this major axis is very small, $\pm$25 \kms.

\subsection{HCO$^{+}$, H$^{13}$CN and vibrationally excited HCN }
\label{s:moment_hcop}

\begin{figure*}[tbh]
\includegraphics[width=6cm, trim = 1.5cm 7cm 0cm 8cm, clip]{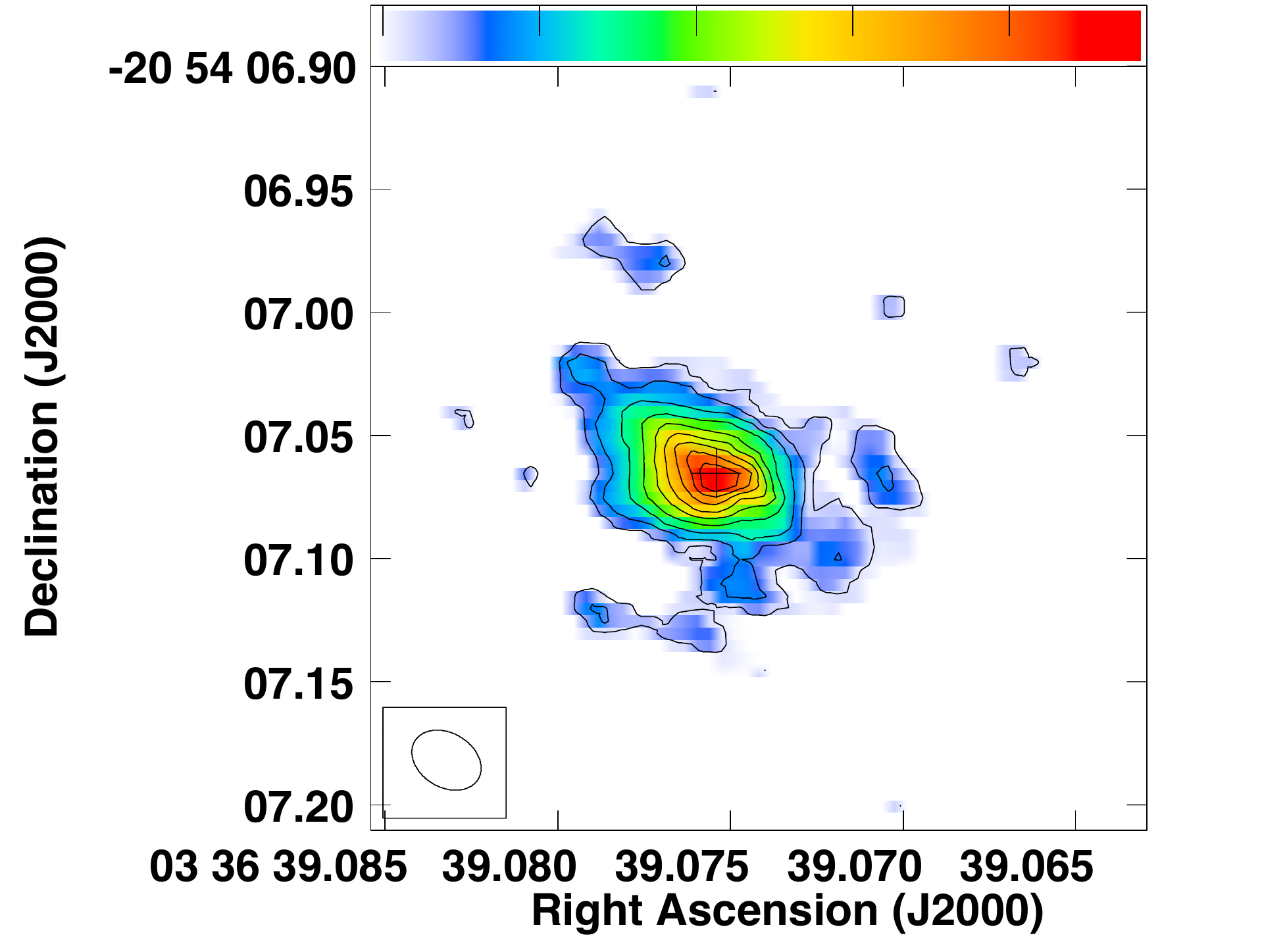}
\includegraphics[width=6cm, trim = 1.5cm 7cm 0cm 8cm, clip]{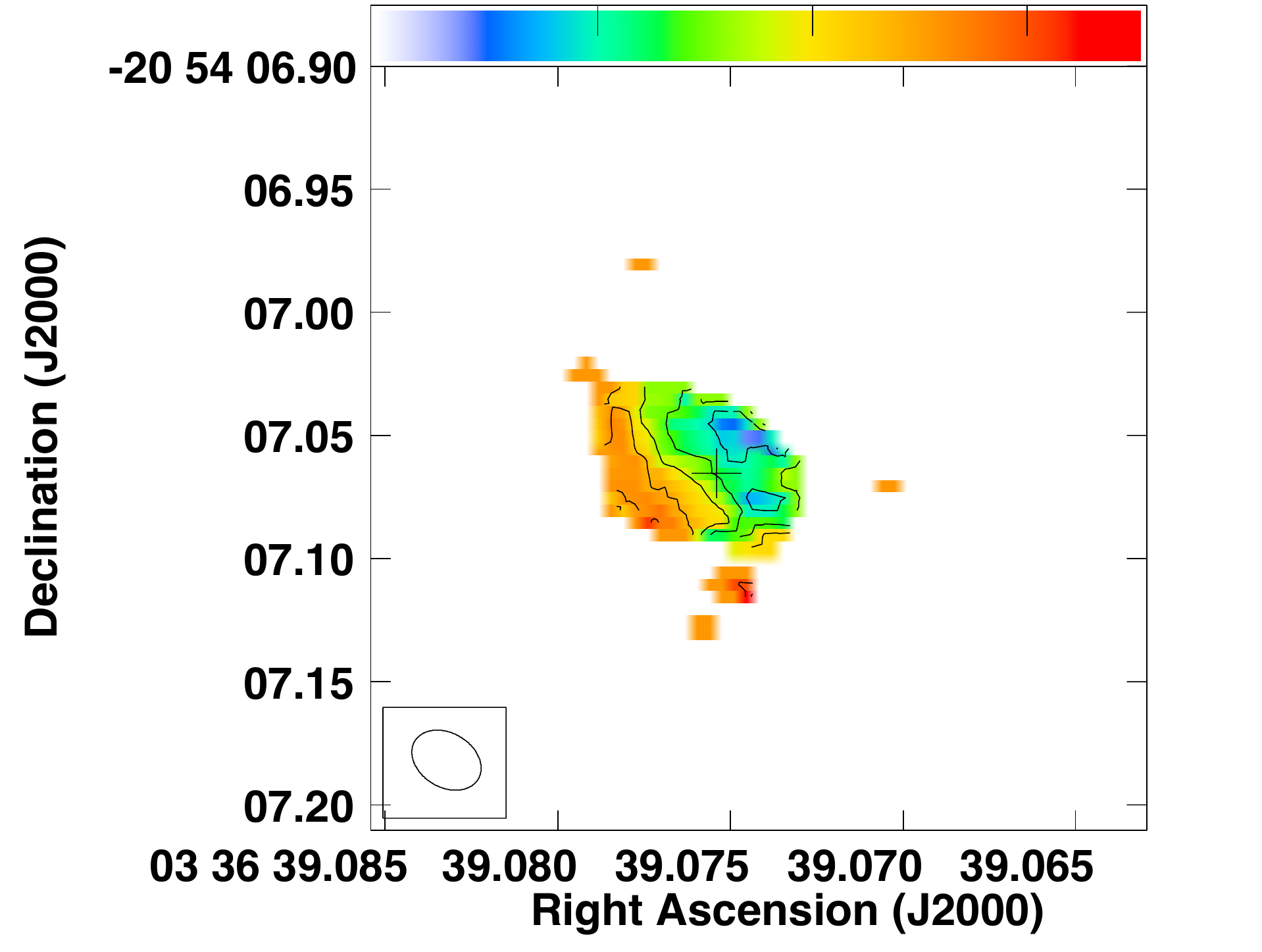}
\includegraphics[width=6cm, trim = 1.5cm 7cm 0cm 8cm, clip]{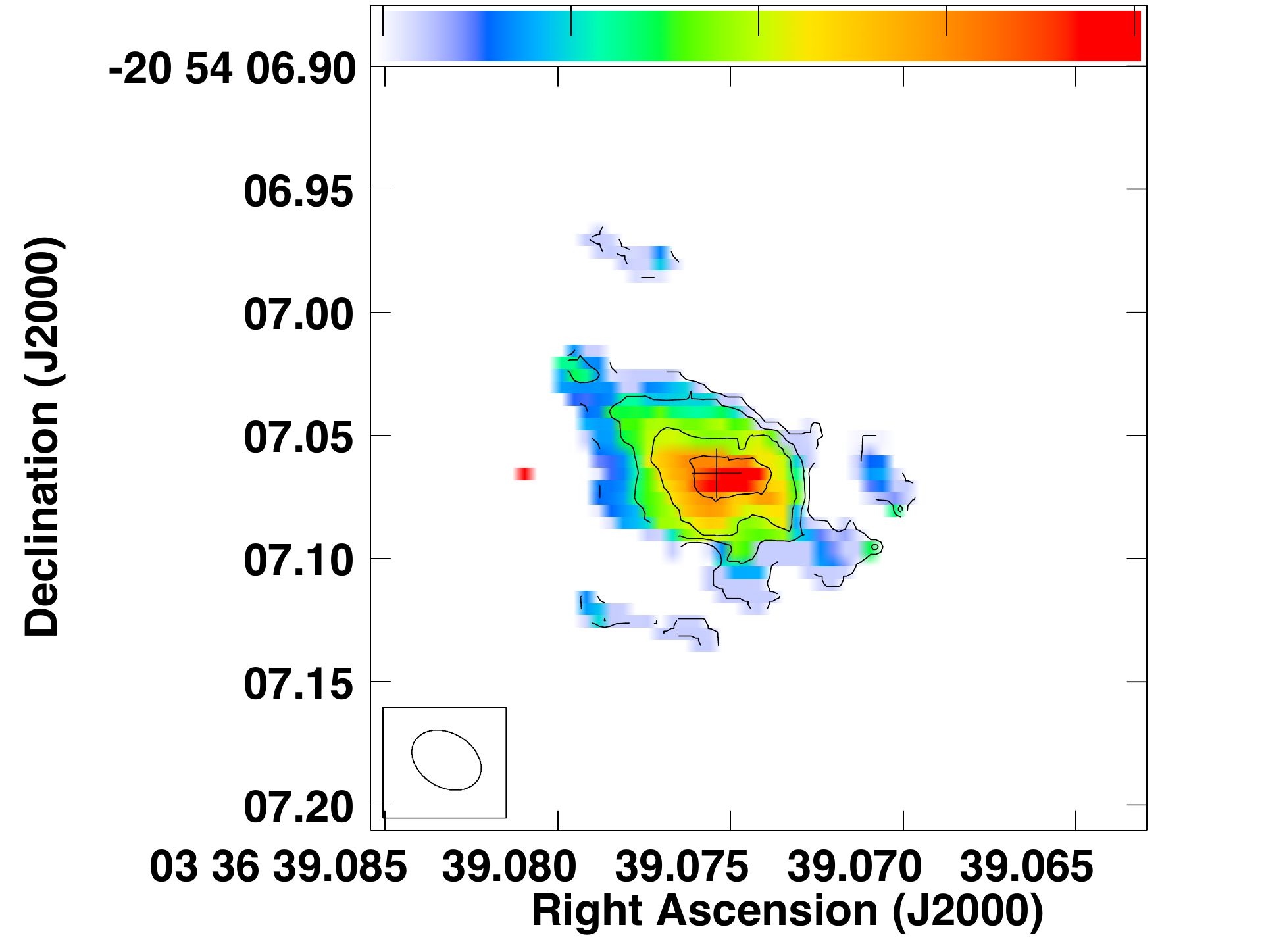}
\caption{\label{f:mom_hcop} HCO$^+$ 4--3 moment maps. Left: Integrated intensity (mom0) where contours are 0.025$\times$ (1, 3,5,7,9,11,13,15,17)  Jy \kms\ beam$^{-1}$. Colours range
from 0 to 0.48 Jy \kms\ beam$^{-1}$. Centre: velocity field (mom1) where contours range from 
1650\kms\ to 1825 \kms\ in steps of 25 \kms, colours range from 1650 to 1825 \kms. Right: Dispersion map (mom2) where contours are 
10$\times$(2,4,6,8) \kms. Colours range from 20 to 100 \kms. The cross indicates the position of the 345 GHz continuum peak.
}
\end{figure*}

 HCO$^+$ 4--3 emission is found (Fig.~\ref{f:mom_hcop}) in a structure with a FWHM size of $4.5 \times 2.5$ pc and a PA of 70$^{\circ} \pm 10^{\circ}$ centred on the nucleus.
 Kinematics are complex and not consistent with a single kinematical component. The velocity field in the centre has a similar PA of 140$^{\circ}$ as for CO 3--2 (Sec.~\ref{s:mom1}).
 However, the intensity weighted moment 1 map is not sensitive to the highest velocity emission, and the velocity field is the result of superposed rotating and non-circular components
 (out- and inflows)(see Sec.~\ref{s:dyn}).  The nuclear spectrum is double peaked with a maximum brightness temperature of 40 K at $v$=1670 \kms.

We also detect  H$^{13}$CN $J$=4--3 and vibrationally excited HCN $J$=4--3 $\nu_2=1f$ ($T = E_{\rm l} / k$=1050 K)(HCN-VIB).
HCN-VIB emission is highly concentrated on the nucleus with a FWHM size of $2.0 \times 1.5$ pc and a PA of 66$^{\circ}\pm10^{\circ}$. The line
is wide with $\Delta V$=350 \kms (FWHM) with a peak $T_{\rm B}$=15 K. We only have blue-shifted  H$^{13}$CN  line emission in our spectral window, but we can see that 
blue-shifted emission is found associated with the jet, but also with the continuum extension to the east of the nucleus (Sec.~\ref{s:continuum}).

We also find lines at redshifted frequency $\nu$=342.26 and 344.5 GHz. The identification of these lines is not clear but we tentatively identify the first as
HC$^{15}$N $J$=4--3 and the second either as vibrationally excited HC$_3$N 
or as SO$_2$. Lower resolution spectra, and a brief
discussion of the line identification in the same spectral set-up, were presented in \citet{aalto16}.

\subsection{Continuum}
\label{s:continuum}

\begin{figure}[tbh]
\centering
\includegraphics[width=9cm, trim = 1.5cm 7cm 0cm 8cm, clip]{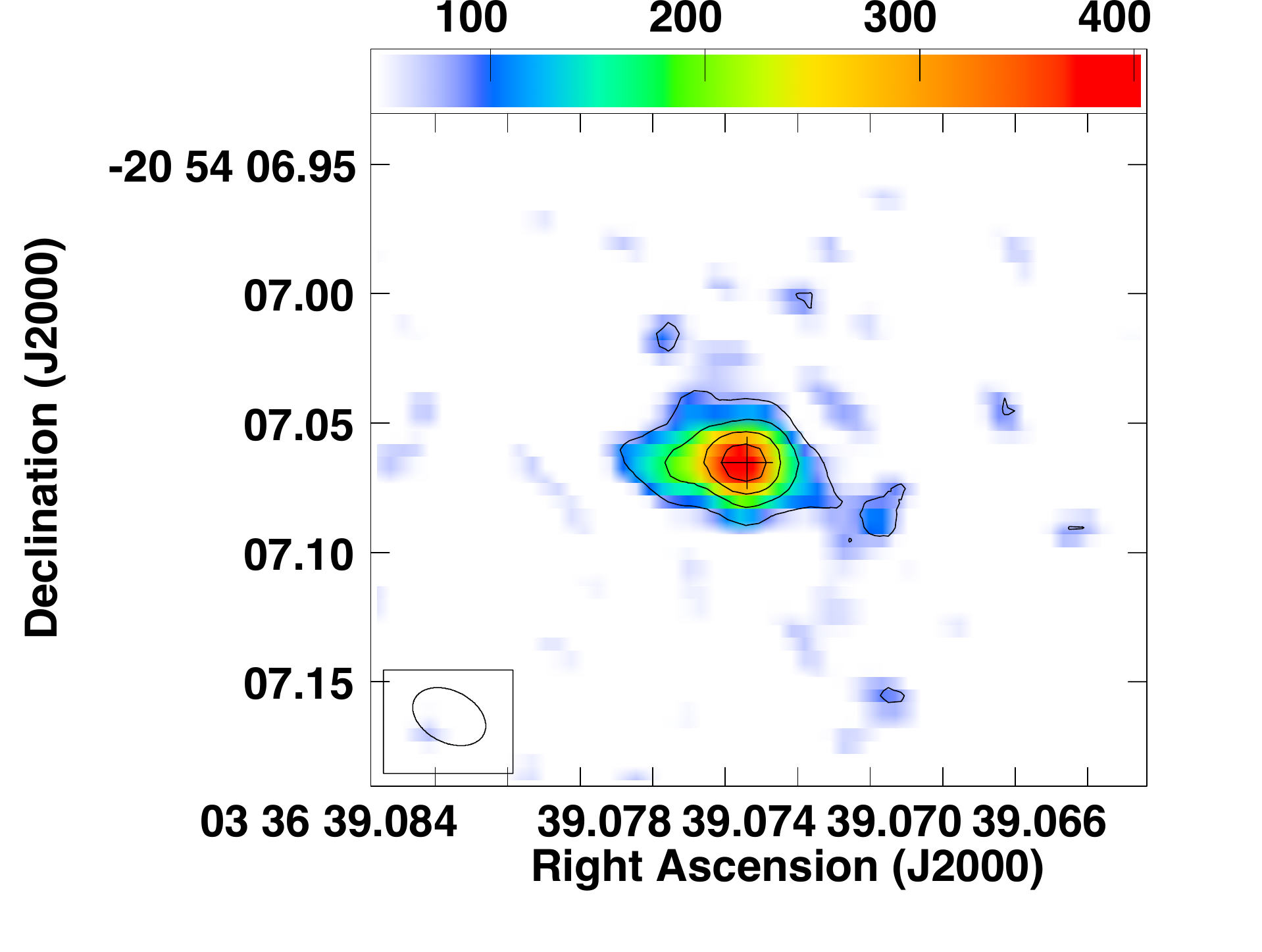}
\caption{\label{f:cont} 0.8~mm continuum (merged 342, 349, 356 GHz line-free channels). Contour levels are 0.085 $\times$(1,2,3,4) mJy beam$^{-1}$ and colours range from 0.05 to 0.4 mJy.  The lowest contour is at 3$\sigma$.
Coordinates are in J(2000) ).
}
\end{figure}

We merged all line-free channels in our observations into a 0.8~mm continuum image (Fig.~\ref{f:cont}) with an rms of 30 $\mu$Jy. The continuum consists of a compact
component and emission extending to the east of the peak. A two-dimensional Gaussian fit gives a FWHM size of  $4.1 \times 2$ pc and a position angle PA=90$\pm 5^{\circ}$. The continuum is faint
(0.41$\pm$0.02 mJy beam$^{-1}$ peak ($T_{\rm B}$=5 K) and 1.3$\pm$0.1 mJy integrated). \citet{aalto16} report a 345 GHz continuum flux of 2.2 mJy. The missing 1 mJy is found in more extended and patchy
 emission, and hence not part of this Gaussian fit. The PA of the high resolution continuum differs from that at lower resolution,  which has a PA of 104$^{\circ}$ \citep{aalto16}. Based on the radio observations of \citet{costagliola16}, we
 estimate an upper limit to the contribution from synchrotron and free-free emission to the 0.8~mm continuum of 3\% and   6\% respectively.

\section{Physical conditions}

\label{s:physical}

\subsection{Gas and dust temperatures}

The high resolution of our ALMA observations allows us to measure meaningful CO 3--2 brightness temperatures ($T_{\rm B}$(CO 3--2)), which can be used as a lower limit to the gas kinetic temperature ($T_{\rm k}$). 
Gas with high $T_{\rm k}$  ($\gapprox 100$ K) is located in the inner $r$=4 pc in a complex structure. The maximum
$T_{\rm B}$(CO 3--2)=180 K is found at a velocity $v$=1700 \kms, blueshifted from $v_{\rm sys}$ and located 0.\asec 02 north of the nuclear continuum peak. The orientation and major/minor axis ratio of the
high-$T_{\rm B}$(CO 3--2) emission is similar to that of the 0.8~mm nuclear continuum (see Sec.~\ref{s:continuum}). Furthermore, the detection of emission from the polar molecule 
HCO$^+$ 4--3 implies that gas volume densities, $n$, are large enough ($n$$>$$10^4$ $\cmmd$) to couple dust and gas (Sec. ~\ref{s:moment_hcop}). We can therefore assume that
$T_{\rm d}$$ \sim $$T_{\rm k}$ and that $T_{\rm d}>$100 K in the central region, and $T_{\rm k}$=$T_{\rm d}$$\sim$180 K in the inner 3 pc.

\smallskip
The  CO 3--2 emission in the jet is clumpy with varying $T_{\rm B}$(CO 3--2). Brightness temperatures reach $T_{\rm B}$(3--2)=40-50 K in the high-velocity gas at $r$$\sim$25-40 pc to the north and south of the nucleus. 
This is also the region where the jet appears to widen before narrowing again further out.  Such elevated gas temperatures in the expansion region may indicate local heating from shocks.

Closer to the nucleus, $r$=10-20 pc,  $T_{\rm B}$(CO 3--2) is 20-25 K in the high-velocity gas while higher values,
$T_{\rm B}$(CO 3--2)$>$50 K, are found for lower velocities in the jet and narrow wind.  Interestingly, the nuclear emission of the more highly excited CO 6--5 emission is associated with the jet (and the northern narrow wind)
while the inner CO 3--2 emission has a higher PA more similar to that of the disk. The difference in uv coverage between the CO 6--5 and 3--2 observations means that comparisons are precarious, but the difference in nuclear
morphology is striking \citep{aalto17}.  Bright, near-systemic ($v$$\sim$1740 \kms), collimated  CO 6--5 emission extends 10 pc to the south of the nucleus, while  this structure is missing both for CO 3--2 and HCO$^+$ 4--3. This
is either because of self-absorption masking emission at $v$=1740 \kms or because the nuclear jet component is hot and dense with elevated CO 6--5/3--2 intensity ratios.  In general  only very little HCO$^+$ 4--3 emission
is associated with the jet (or wind) outside the nuclear launch regions.

In the gas aligned with the stellar disk, warm ($T_{\rm B}$(CO 3--2)$\sim$50 K) gas is found in the inner 0.\asec 2. Further out values drop to 20-30 K.

\subsection{Velocity dispersion and turbulence}

\begin{figure}[tbh]
\includegraphics[width=9cm, trim = 1.1cm 4cm 0cm 4cm, clip]{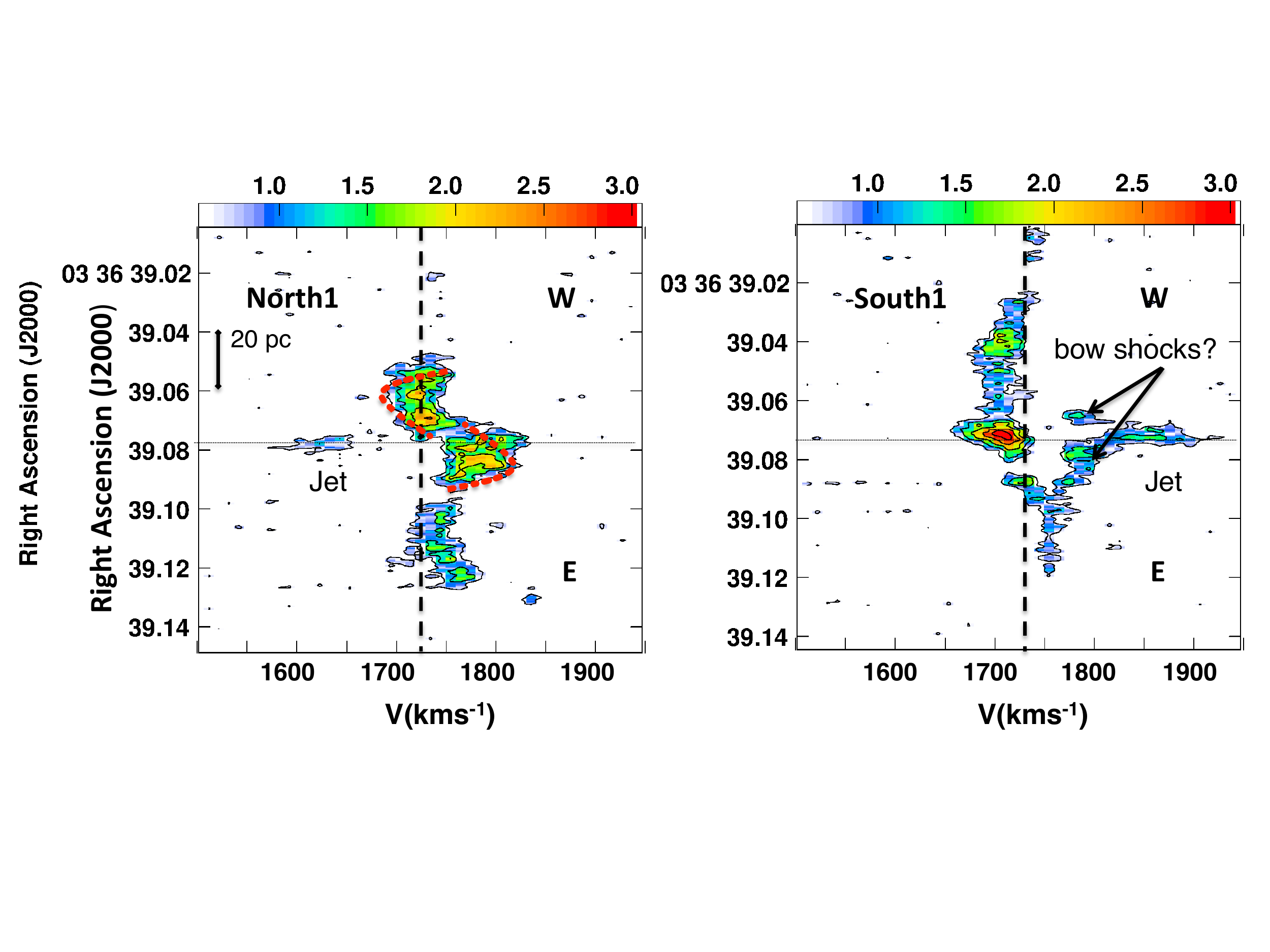}
\includegraphics[width=9cm, trim = 1.1cm 4cm 0cm 4cm, clip]{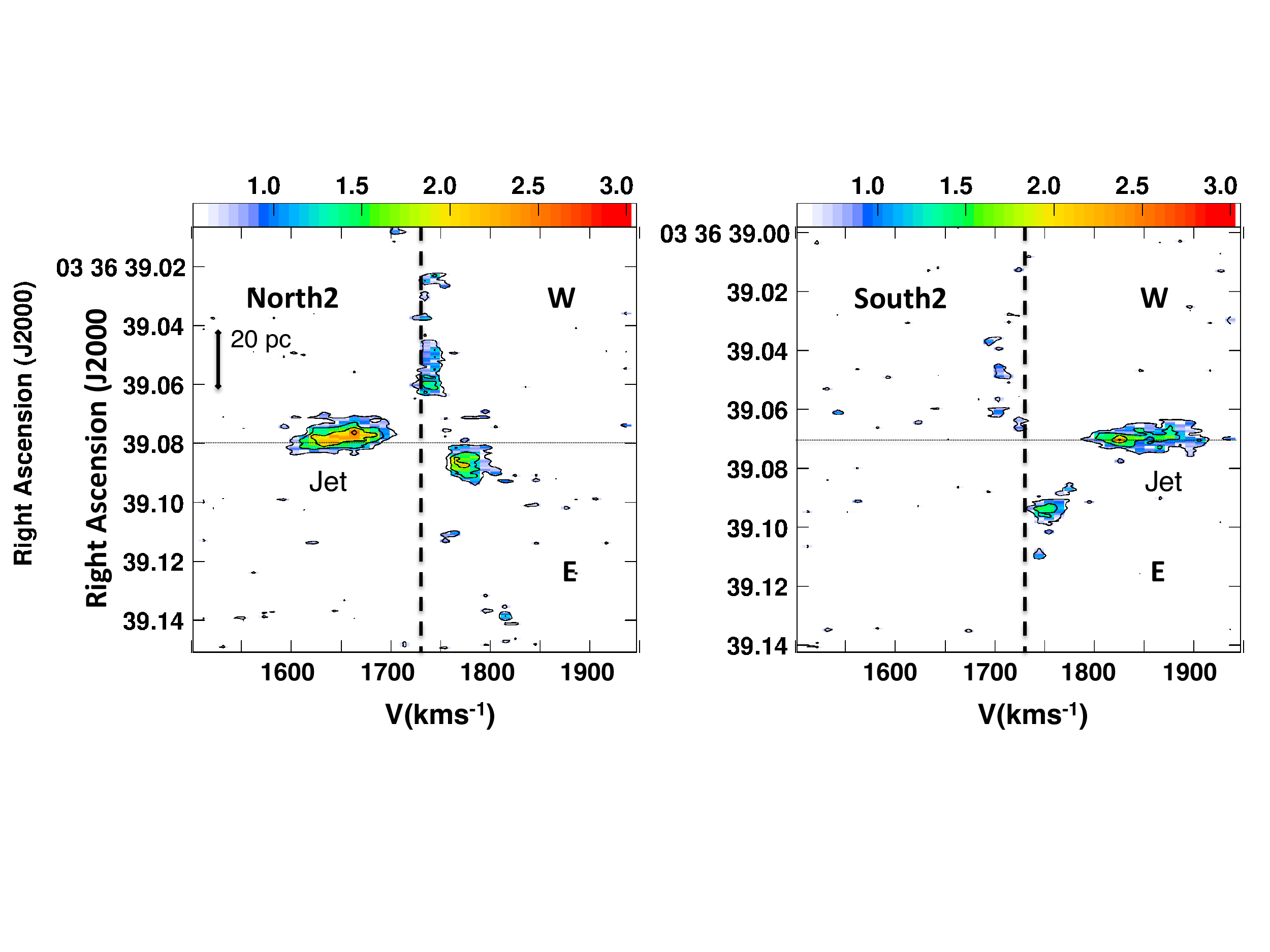}
\caption{\label{f:jetcut}  Position-velocity (pV) diagrams of CO 3--2 across the jet axis. Contour levels are 0.6$\times$(1,2,3,4) mJy beam$^{-1}$, and colour range from 0.6 to 30 mJy. 
The horizontal dotted line indicates the jet symmetry axis. Top left panel: cut 0.\asec 15 to the north of the nucleus. The expected signature of a rotating wind is indicated with red dashed
curves (see Fig.~\ref{f:wind_model} and Sec.~\ref{s:wind_model}). Lower left panel: cut 0.\asec 3 to the north of the nucleus.
Top right panel: cut 0.\asec 15 to the south of the nucleus; Lower right panel: cut 0.\asec 3 to the north of the nucleus. 
The arrows point to structures that are proposed to be bow shocks (see Sec.~\ref{s:bow}). 
}
\end{figure}

High values of $\sigma$ in the jet (Fig.~\ref{f:mom2}) are due to emission with high intrinsic velocity width and, in some locations, also to overlapping multiple, narrower emission features along the line-of-sight.  
The elevated intrinsic dispersion ($\sigma$=40-60 \kms), is found from the nucleus out to the jet expansion at $r$$\sim$25-40 pc. Position-velocity (pV) diagrams
across the jet axis  (Fig.~\ref{f:jetcut}) show that $\sigma$ in the wind (narrow and slow) is lower than that in the jet. Some of the large line widths in the jet may stem from
expansion or unresolved rotation. However, velocity dispersions in the jet and wind are still very high by comparison to normal giant molecular clouds  or cloud cores on similar size scales. Therefore, standard
CO luminosity to H$_2$ mass conversion factors are unlikely to apply (see Sec.~\ref{s:energetics}).  The turbulent jet and wind of NGC~1377 are also unlikely sites of ongoing star formation, a possible scenario 
suggested for other galactic-scale outflows \citep{maiolino17}. 

The origin of the large line widths may be internal working surfaces from variable jet emission or linked to jet driving processes (see Sec.~\ref{s:bow}).

\section{Nuclear gas and dust properties}

\label{s:nucleus}

\begin{figure}[tbh]
\centering
\includegraphics[width=9cm, trim = 4cm 4cm 3cm 4cm, clip]{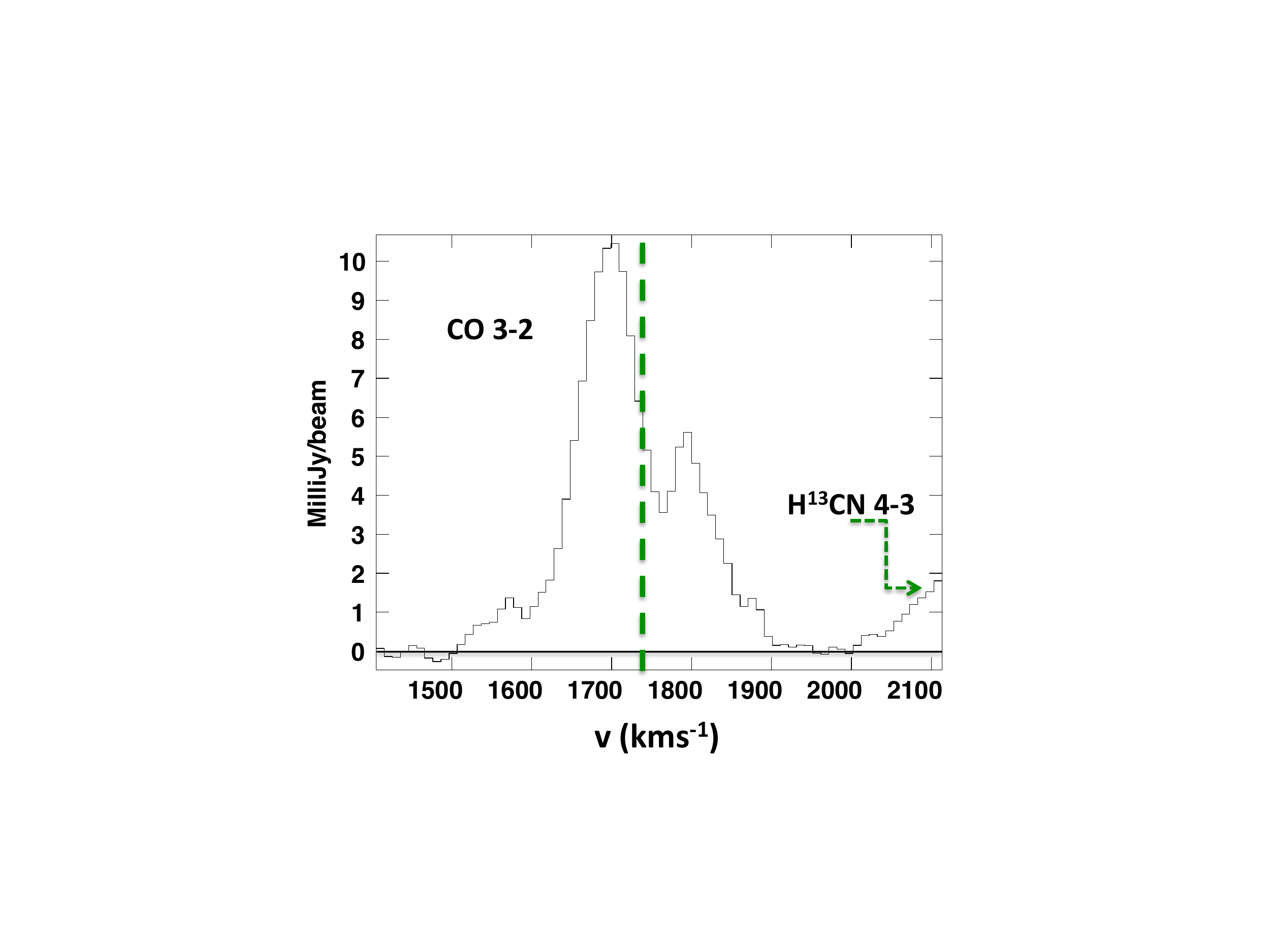}
\caption{\label{f:spec}  Nuclear spectrum of NGC~1377 showing the asymmetric line profile of CO 3--2  (there
is an  H$^{13}$CN 4--3 feature to the right).  The green dashed line indicates systemic velocity.
}
\end{figure}

\begin{figure}[tbh]
\centering
\includegraphics[width=9cm, trim = 2cm 0cm 0cm 0cm, clip]{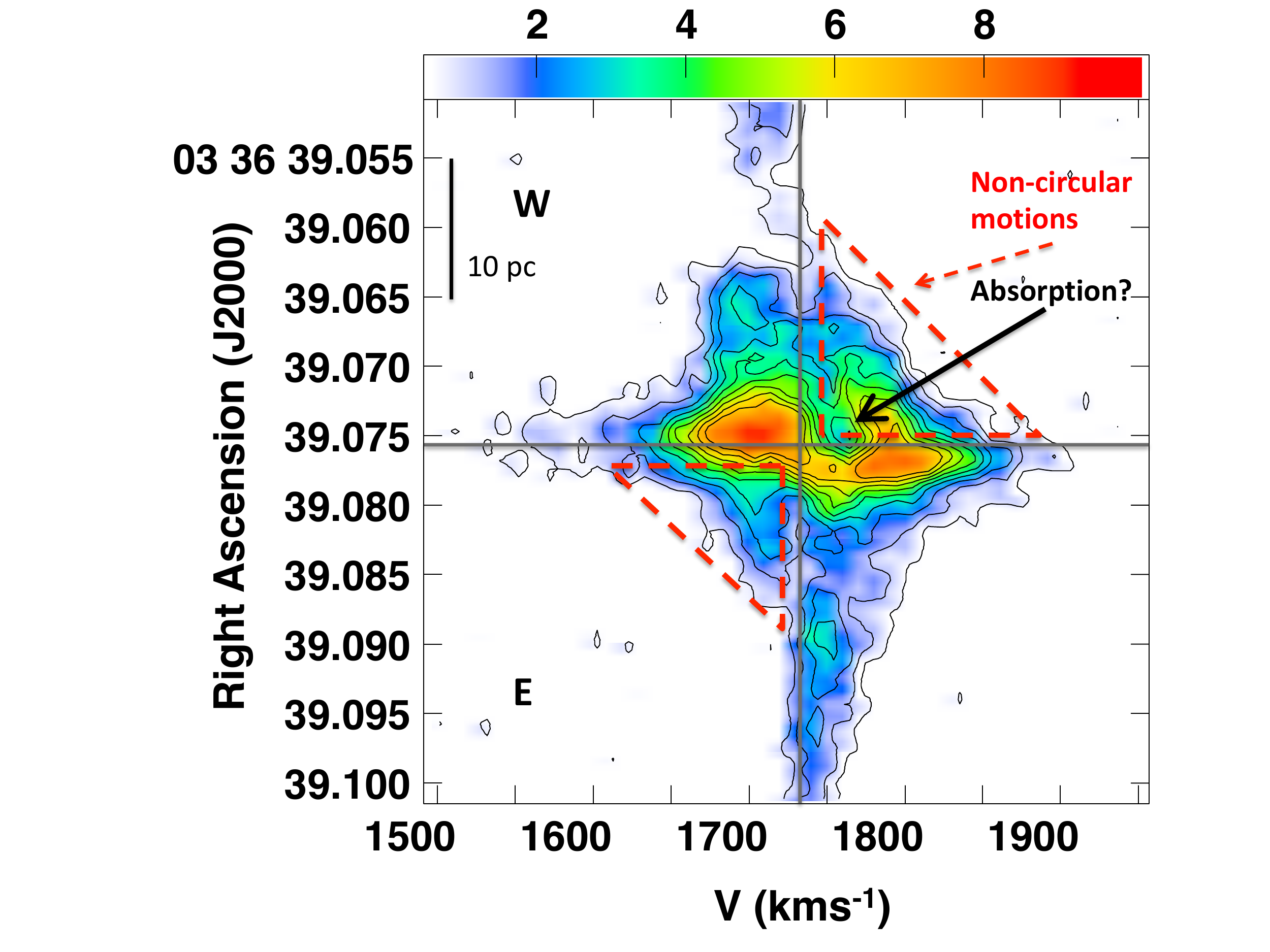}
\caption{\label{f:pv-nuc} CO 3--2 pV diagram cut across PA=90$^{\circ}$ along the nuclear major axis. Red dashed triangles show regions of non-circular velocities. The arrow indicates
the potential absorption structure on the nucleus and to the west. 
}
\end{figure}

\subsection{Nuclear morphology, dynamics, and enclosed mass}
\label{s:dyn}

The SMBH of NGC~1377 appears obscured by an asymmetric disk or torus of dimensions $r$$\simeq$4 pc for CO 3--2, $r$$\simeq$2 pc for HCO$^+$, and $r$$\simeq$1 pc for HCN-VIB. The
presence of a nuclear warp is supported by changing position angles (PA) on different size scales. From a PA of 104$^{\circ}$ for the 0.8~mm continuum on scales of 20 pc \citep{aalto16} to
PA=90$^{\circ}$ on scales of 4 pc.
The difference in orientation of the launch region between the jet and the narrow wind, also supports the scenario of a complex nuclear warp. Furthermore, the lower PA of the HCN-VIB emission 
(excited by  high surface brightness mid-IR dust emission \citep{sakamoto10,aalto15}) implies that polar hot dust may also contribute to the obscuration of the nucleus. This has also been suggested
for other galaxies, for example the nuclear obscuration in the Seyfert galaxy Circinus \citep[e.g.][]{tristram14,izumi18}.

The dimension and physical conditions of the NGC~1377 torus is similar to those of other torii imaged at high resolution so far \citep[e.g.][]{garcia16,impellizzeri19,garcia19, combes19}. The H$_2$
column density does however appear to be higher in NGC~1377 on similar scales: The core component of the 0.8~mm continuum has a brightness temperature of $\sim$5 K. For $T_{\rm d}$=180 K
the dust opacity at $\lambda$=0.8~mm is $\tau$$\simeq$0.04.  Using the formalism of \citet{keene82} $N$(H + H$_2$)/$\tau_{\lambda}$=$1.2 \times 10^{25}(\lambda/400 \mu{\rm m})^2$ $\cmmt$,
we find a $N$(H + H$_2$)$\simeq$$1.8 \times 10^{24}$ $\cmmt$. This suggests that there might be a Compton Thick (CT) dust structure with radius $r$$\simeq$1.5 pc around the NGC~1377 nucleus.
The column density is higher than that previously estimated from the 690~GHz continuum for a larger (5 pc) beam \citep{aalto17}, but is consistent for the smaller radius found for the 0.8~mm continuum.
However, if the dust is actually hotter than $T_{\rm d}$=180 K, then $1.8 \times 10^{24}$ $\cmmt$ is an upper limit to $N$(H + H$_2$). We exclude contributions from synchrotron and free-free to the 
0.8~mm continuum emission due to the extreme radio-quietness of NGC~1377 \citep{costagliola16}, but this requires further investigation.

The nuclear high-velocity gas appears to be in Keplerian rotation with a PA of 100$^{\circ}$ (Sec.~\ref{s:highest_vel}).  The actual inclination of the nuclear gas is difficult to determine and is likely changing
with radius due to the suggested presence of a warp.  The high resolution 0.8~mm continuum implies a nuclear inclination $i\gapprox$60$^{\circ}$ suggesting a dynamical mass $M_{\rm dyn}$ of $0.9 \times 10^7$ \msun. 
To this we added the uncertainty in position of the nuclear high-velocity gas discussed in Sec.~\ref{s:highest_vel}, and the possibility that the inclination could be higher,  which results in $M_{\rm dyn}$=$9^{+2}_{-3} \times 10^6$ \msun.

This $M_{\rm dyn}$ is higher than previously estimated enclosed masses for NGC~1377 \citep{aalto12b,aalto16,aalto17} and we attribute this to the higher
sensitivity and resolution of the new CO 3--2 observations. Rotational velocities can be separated from non-circular motions to a higher degree than before.  At such small radius $M_{\rm dyn}$ is expected to be dominated by
the SMBH and the high mass is interesting since it is above the value suggested by the  $M-\sigma$ relation \citep[e.g.][]{graham11}. This is discussed further in Sec.~\ref{s:growth}.

An additional blue-shifted high-velocity component is seen on the red side of nuclear rotation (Fig~\ref{f:rot}). If real, this may be an in- or outflowing gas component, or it is linked to a second nucleus. A binary SMBH is a 
tantalizing scenario in the context of NGC~1377 as a merger remnant.  Two SMBHs could drive jet-precession or, with different inclinations of the disks, do away with the need for jet precession to explain the velocity reversals.
Further studies at higher resolution are necessary to further investigate the possibility of a binary SMBH.

The PA of $100^{\circ}$  for the nuclear dynamics is different from the previously suggested orientation of nuclear rotation. In  \citet{aalto17} the CO 6--5 high-velocity gas has an apparent PA of
$140^{\circ} \pm 20^{\circ}$, but the CO 6--5 velocities are a factor  of two lower than the high-velocity gas traced with our new CO 3--2 data. We do find this dynamical component (70 - 80 \kms) in the moment 1 maps of
CO 3--2 and HCO$^+$ 4--3.  We propose that this intermediate velocity gas is part of the wind and jet rather than the nuclear disk (see Sec.~\ref{s:origin}).

\subsection{Indications of inflowing gas}
\label{s:flow}

The CO (Fig.~\ref{f:spec}) and HCO$^{+}$ spectra are double-peaked in the centre, with maxima at velocities: $\sim$1670-1700 \kms and $\sim$1800-1810 \kms. CO 3--2 and HCO$^{+}$ 4--3
emission avoid the location of the dust continuum peak at velocities near systemic. The continuum is too faint to cause the depression of the systemic CO and HCO$^{+}$ emission,
which is either the result of a dynamical structure or self-absorption in lower-excitation foreground gas. The latter is a more likely scenario since the double peak is not seen
in HCN-VIB (probing high column density hot gas \citep[e.g.][]{aalto15b}) and the dust continuum peak is also consistent with large gas column densities here. 

The (potential) absorption depth peaks at velocity 1760 \kms, $\sim$20-30 \kms redshifted of systemic. It may indicate that lower excitation foreground gas is moving in towards the centre of NGC~1377, possibly
originating in the envelope of gas that may be fed by the slow wind \citep[e.g.][]{evans99} (see also  (Sec.~\ref{s:bow})). The  position velocity (pV) diagram  (Fig.~\ref{f:pv-nuc})  shows the purported absorption feature
occurring mostly west of the nucleus, on the "forbidden" side of non-circular motions. However, we caution that the interpretation of the dynamical origin of gas in absorption features is complex.
 
The pV diagram also reveals emission at non-circular velocities, starting 10-15 pc from the nucleus on both sides of the centre, although there is an asymmetry with stronger emission on the western
side. This emission may also be attributed to fast inflowing gas. Models of rotating inflows predict similar velocities to the Keplerian at these distances from the central object, although the radial dependence is different
than Keplerian \citep[e.g.][]{oya14}. However, given the combined presence of a  nuclear warp and disk-wind it is likely that we see overlapping structures of inflow and outflow, that are difficult to disentangle. 

The gravitational (Bondi) radius of influence of a $9 \times 10^6$ \msun\ SMBH (with stellar velocity dispersion $\sigma$=83 \kms\ \citep{aalto12b}) is $r_{\rm g}$$\simeq$4 pc. \citet{hopkins12} suggest that near $r_{\rm g}$, 
systems become unstable to the formation of lopsided, eccentric ($m$ = 1 mode), precessing gas+stellar discs. Strong torques induces shocks and inflows, which may in turn help drive outflows. The dimensions
of the nuclear disk in NGC~1377 are similar to the expected size of the SMBH radius of influence.

\subsection{Accretion luminosity or compact star formation}
\label{s:ex}

To estimate the luminosity of the nuclear dust structure we adopt the same method as \citet{aalto17}. Fitting the luminosity to a simple spherical dust structure with the average radius of $r$=1.5 pc and $T_{\rm d}$=180 K results in $L_{\rm core}$$\simeq$$4.8 \times 10^9$ \lsun\footnote{This is a factor of 4 higher than the relatively larger, but cooler nuclear dust structure suggested in \citet{aalto17}.}

If the luminosity is emerging from an embedded starburst, an ensemble equivalent to $\sim$$5 \times 10^3$ O-stars (with a luminosity of $10^6$ \lsun\ each) could produce the luminosity. The approximate total mass of such an ensemble
of O-stars is $5 \times 10^5$ \msun\ and  for a normal Salpeter initial mass function (IMF) the mass in low mass stars would be ten times that of the O-stars, resulting in a total mass in starburst stars of $5.5 \times 10^6$ \msun.  This is
within the allowed range of the dynamical mass\footnote{We are excluding a contribution from a pre-existing nuclear stellar cluster (NSC) due to the small dimensions of the nuclear dust structure.}.

However, to bury the emission from the stars of a starburst in the obscuring disk would require a very small radius ($r<0.6$ pc to allow for enough obscuration)  of the stellar distribution. This 
is much smaller than typical sizes of stellar clusters of this mass (2-10 pc) and the stars have to be packed extremely close. In addition,  the spectrum of the faint radio emission detected has a synchrotron spectrum \citep{costagliola16}
which does not suggest free-free emission from a hot plasma irradiated by massive young stars.

\smallskip
If the luminosity is instead emerging from the accretion onto an SMBH, there is no need to add a young starburst population to the nuclear mass, and $M_{\rm SMBH}$=$9  \times 10^6$ \msun. The Eddington luminosity
of the SMBH is approximately $2.7 \times 10^{11}$ \lsun\ and the growth of the NGC~1377 black hole would occurs at a rate of $\sim$2\% Eddington to produce the luminosity.


\section{Energetics and turbulence of the jet and wind}
\label{s:energetics}

\subsection{Molecular masses}
Adopting a standard, Galactic  CO-to-$H_2$ conversion factor, X(CO) the molecular mass in the jet is estimated to $2.3 \times 10^7$ \msun\ \citep{aalto16}\footnote{for gas moving at higher (projected) velocities
than 60 \kms\ which would include $>$$70$\% of the volume if the jet is smoothly precessing.}.   However,  \citet{aalto15a} found that a Galactic
X(CO) overestimates the mass in turbulent outflowing gas by a factor $\sim$5. Applying this factor, we instead find a jet gas-mass of $M_{\rm jet}$$ \simeq$$5 \times 10^6$ \msun\ (not corrected for any missing emission
at lower velocities). For the narrow wind (NW) we estimate a mass of $M_{\rm NW}$$ \simeq$$8 \times 10^6$ \msun, also by applying the lower X(CO) for turbulent gas. Half of the flux of the slow wind (also referred to as 
the envelope or cocoon (Sec.~\ref{s:bow})) (SW/C) emerges from outside the turbulent region and we estimate its mass to $M_{\rm SW/C}$$ \simeq$$1.8 \times 10^7$ \msun. Also in the disk about 50\% of the emission is turbulent, resulting in 
$M_{\rm disk}$$ \simeq$$5.5 \times 10^7$ \msun. These numbers have significant uncertainties and a study of the size-linewidth relation for the clouds in NGC~1377, should provide a better
estimate of the cloud stability spectrum and the X(CO) of the various phases of the molecular gas in NGC~1377.

\subsection{Outflow velocities} 
\subsubsection{Wind velocity}

Velocities found in the narrow wind can be $>$80 \kms\ above systemic (a combination of projected rotation and outflow motions). Comparing north and south average wind velocities, a projected outflow speed of the
narrow wind is $\sim$30 \kms. Assuming an average inclination $i$=70$^{\circ}$ of the launch region, the $v_{\rm out}$ of the narrow
wind is  90$ \pm 30$ \kms\ (uncertainties stem from inclination, opening angle and that $v_{\rm out}$ varies with radius). The slow wind  has projected outflow velocities in the range of 0-20 \kms\ ($v_{\rm out}$=0-60 \kms).

\subsubsection{Outflow velocity of a precessing jet}
\label{s:jet-flip}

\begin{figure}[tbh]
\centering
\includegraphics[width=9cm, trim = 0cm 0.3cm 10cm 9cm, clip]{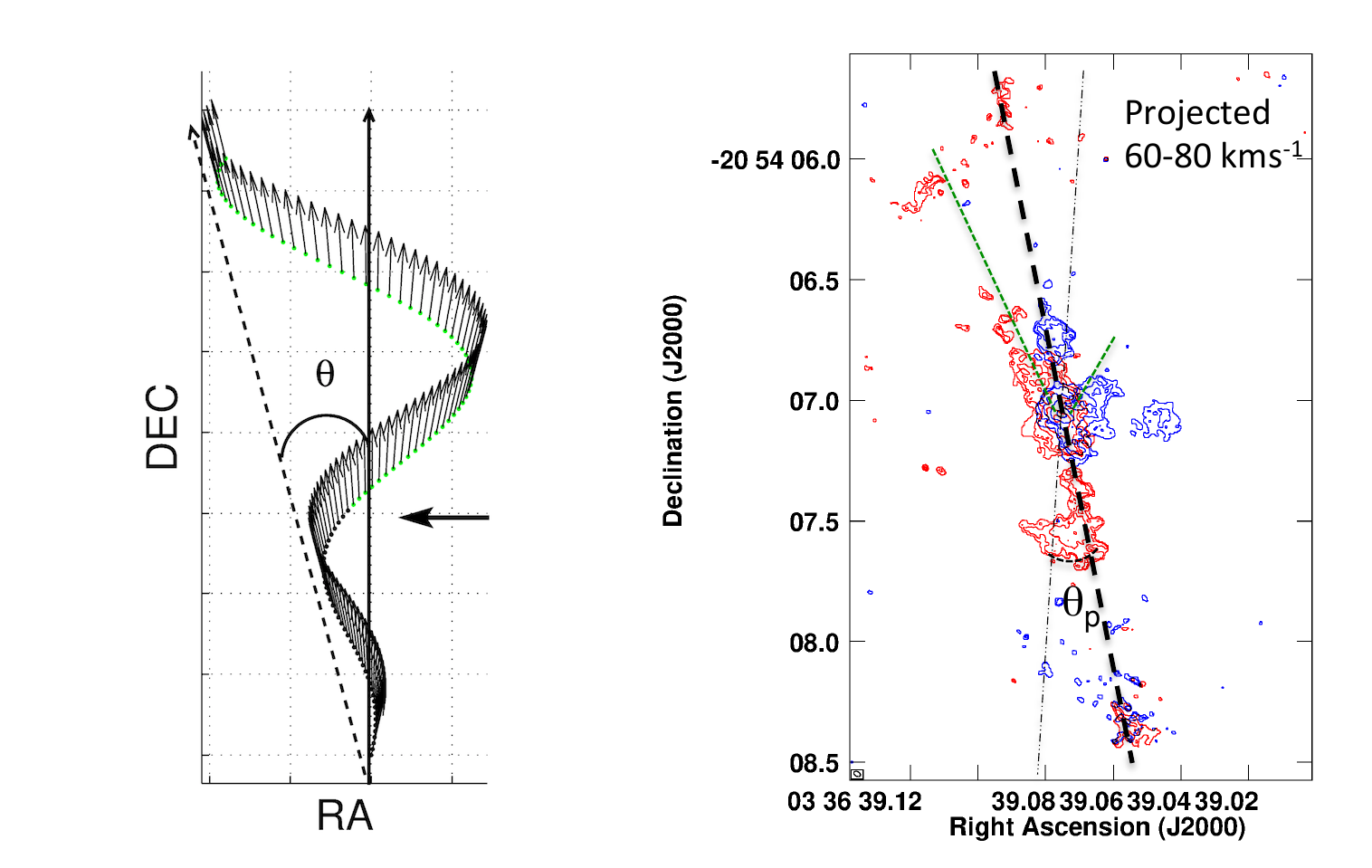}
\caption{\label{f:jet_angle}  Left: Schematic of the simple precession model by  \citet{aalto16} showing the precession angle $\theta$. Right: Lower velocity
(60-80 \kms) jet emission (from Fig.~\ref{f:jet_channel}). Contour levels are (for the blue- and redshifted emission respectively)
 0.004$\times$(1,5,10,20,50) Jy beam$^{-1}$ kms$^{-1}$. The thick black dashed line shows the orientation of the jet axis. The narrow wind is outlines in green and the suggested precession angle in
 a thin, dashed line }
\end{figure}

The jet velocity switches from blue-to-red to the north, and vice versa to the south, which were previously interpreted as evidence of jet precession by \citet{aalto16}. A simple model 
with a precession angle $\theta_{\rm p}$=10$^{\circ}$-25$^{\circ}$ and jet outflow velocities $v_{\rm out}$=240 to 850 \kms\ can reproduce observations. The outflow velocity estimates are highly
dependent on $\theta_{\rm p}$ and the inclination of the jet with respect to the plane of the sky. As pointed out by \citet{aalto16}, the north and south are relatively symmetric in velocity which suggests 
that the jet symmetry axis is close to the plane of the sky. Deviations from the jet symmetry axis are apparent, but determination of $\theta_{\rm p}$ is difficult since the lower velocity jet emission
blends in with the slow wind, and jet-wind interactions distort the symmetry. In addition, the jet emission appears patchy and is likely episodic.  However, an assessment of off-symmetry jet-emission
in Fig.~\ref{f:jet_angle},  gives $\theta_{\rm p}$$\simeq$15$^{\circ}$. For this $\theta_{\rm p}$ the jet may be a maximum of 2$^{\circ}$ away from the plane of the sky to fit the observed velocity reversals.  
A  $\theta_{\rm p}$=15$^{\circ}$ implies a jet $v_{\rm out}$ of $\sim$600 \kms\ ($\pm$60 \kms to account for uncertainties in the projected velocity). In this simple scenario it is assumed that
the jet velocity is dominated by outflowing motions and that the element of rotation is small (Sec.~\ref{s:jet_driving}). The jet emission is changing direction with time, but it is possible that these
variations are not caused by a smooth "swirling" of the jet, but by more sudden, directional changes caused by an uneven accretion flow. The driving force behind the jet precession will be discussed in
a future study.

\subsection{Kinetic and gravitational energy}
\label{s:energy}

In Table~\ref{t:energy} we list approximate estimates of the systems kinetic and gravitational energy (on scales of $r$$\sim$100 pc) together with the velocities for the various components. 
If the gravitational energy of the system is large enough, the rotating and infalling material could be converted to the outflow energy aided by a magnetic field.  The kinetic energy is dominated
by the jet and the gravitational energy by the stellar and gas envelope component. There is rough equipartition between the two, which may indicate a dynamical link. The driving of the jet and wind is discussed in
Sec.~\ref{s:origin}.

The turbulence is high throughout the inner 50-60 pc of NGC~1377 and there is enough kinetic energy in the jet to drive the turbulence in the inner part of NGC~1377 (for jet $v_{\rm out}$ down to 120 \kms). 
A potential link between the jet and its surrounding is discussed in Sec.~\ref{s:bow}.

{\bf
\begin{table}
\caption{\label{t:energy} {Kinetic and gravitational energy}}
\begin{tabular}{lccccc}
 & \\
\hline
\hline \\ 
Component & Velocity & $\sigma$ &  $K^a$  & $W^a$ \\
 & (\kms) &  (\kms) &    & \\
\hline \\
Jet & 600 & 30-60 &   11.2  & 0.01 \\
NW & 90 & 30 &  0.4  & 0.03  \\
SW/C & 30 & 10 &   0.2 & 0.03 \\
Disk &  27 & 10-30   & 0.3  & 0.80 \\ 
BH &   &  &     & 0.03 \\ 
Stars &  &    &  & 17 \\ 

Total$^b$  &  & & $\sim$12 & $\sim$18  \\
\\
\hline \\

\end{tabular} 

$^a$ The kinetic energy is $K$$ \simeq$$\frac{1}{2}M(v^2+ \sigma^2)$ where $M$ is the mass of the component (jet, NW=narrow wind, SW/C=slow wind (also the envelope or cocoon), disk), 
$v$ its velocity and $\sigma$ its dispersion. In units of $\times 10^{11}$ \msun\ $({\rm pc/Myr})^2$. For the SW/C we assume an average outflow velocity of 30 \kms. The rotational velocity of the disk
is assuming an $i$=70$^{\circ}$. The gravitational energy is
$W \simeq G \frac{M^2}{r}$ where $M$ is the
mass of the component.  Units are $10^{11}$ \msun\ $({\rm pc/Myr})^2$. $W_{BH}$ is the
gravitational energy of the central SMBH inside $r_{env}$=100 pc and $W_{star}$ is that of the stellar and envelope system for the same region. The mass of the stars inside $r$=100 pc
is estimated to $2 \times 10^8$ \msun\ from an HST H-band image (HST program GO14728,  J Gallagher PI). 

$^b$ The total kinetic energy of the system $K_{\rm total}$=$K_{\rm jet}+K_{\rm wind}+K_{\rm disk}$ and the total gravitational energy of the system
$W_{total}$=$W_{\rm jet}+W_{\rm env}+W_{\rm disk} + W_{\rm BH} + W_{\rm star}$ out to a radius $r$=100 pc.


\end{table}

}

\subsection{Bow shocks and jet-wind interactions}
\label{s:bow}

Simulations of jets with directional changes, and with varying jet velocity, show that the jet has transverse extensions arising from the presence of bow shock wings trailing behind each internal working surface. This produces
laterally extended emission of lower radial velocities than the jet beam \citep[e.g.][]{raga01}. This can be seen in the NGC~1377 molecular jet in several locations: In Fig.~\ref{f:jet_channel} (the panel with projected velocities
80-110 \kms) curved, blue-shifted  bow-shock structures are particularly apparent  1\asec to the south, and a prominent redshifted curved structure at 0.\asec 5. In the lower velocity panel (projected velocities 60-80 \kms) these
curved structures widen and it is more clear that they are predominantly found on the eastern side of the jet axis.  \citet{tafalla17} observed and modelled velocity gradients across bow-shock regions in terms of expanding "disks"
of laterally ejected material. Such a velocity shift is apparent in the northern widening of the jet emission (Fig.~\ref{f:mom2}. (There are also smaller scale wiggles in the jet structure
(seen in the first portion of the jet out to about 20 pc) that are possible results of helical kink instabilities in the jet \citep[e.g.][]{yasushi93}. )

In the pV diagrams across the jet (Fig.~\ref{f:jetcut} (in particular panel "South1"))  the expected signature of the trailing, extended and lower velocity 
bow shock wings can be seen. The signs of bow shocks and the knotty, clumpy appearance of the molecular jet are possible signatures of internal working surfaces. Furthermore, working surfaces also arise from the
interaction between jet and wind material, which is particularly complex and broad if the jet is changing direction. The more prominent bow-shock structures to the south-east may be due to the presence of more molecular
material here, compared to the corresponding north-western side which appears vacated of gas (see e.g. Fig.~\ref{f:jet_sys}).  \citet{tabone18} discuss
jet-wind interactions and how a disk-wind may refill v-shaped regions emptied by jet bow-shocks. Located north-west of the jet may be a such a bow-shock cavity which the disk-wind
currently has not filled. The jet cocoon (left behind by the leading working surface) produces a low-velocity emission component  that can be very extended. It is possible that the slow, wide-angle emission component that surrounds
the jet and narrow wind is such a cocoon.


\section{Origin and driving of wind and jet}
\label{s:origin}

\subsection{Driving mechanism of the narrow wind - a potential rotating magneto-centrifugal wind.}
\label{s:wind_driving}

In \citet{aalto16} a scenario where the jet is entraining and accelerating a very slow, wide-angle minor axis molecular outflow was preferred over a direct-driven wind. Here, with our  new high resolution data, we
can resolve the dynamics of the narrow wind and also trace it back to the nuclear disk while the slow wind is more extended and diffuse.  In contrast to our previous conclusion, we now suggest that the
narrow wind is primarily direct driven and that it is launched from the nuclear disk as a rotating disk-wind.  

The narrow wind shows an east-west velocity shift (Sec.~\ref{s:narrow})  which is consistent with the rotation of the nuclear disk.  Winds and outflows can be magneto-centrifugally driven, powered by rotation and
gravitational energy and launched along magnetic field lines. In  this  scenario,  a  large-scale poloidal  magnetic  field  threads  the  disk  in  the  vertical  direction. When they are driven from the inner edge of the disk
they are often referred to as X-winds \citep[e.g.][]{shu95}, and when they are launched from further out (involving a larger section of the disk) they can be referred to as a disk-wind \citep[e.g.][]{pudritz05}. Magnetohydrodynamic
(MHD) outflows are launched from rotating sources, the rotation of the jet and outflow is an expected feature of MHD jet and outflow formation. If a wind is emerging from the centre of a rotating disk, the outflow material is
ejected with a significant amount
of angular momentum, which may become conserved along the streamlines. In contrast to thermal winds, for example, MHD winds not only remove mass from the disk, but also exert a torque and remove angular
momentum from the disk. 

There needs to be sufficient free electrons mixed in with the molecular gas to anchor the magnetic field. The presence of emission from molecular ions (HCO$^+$) in the inner region of NGC~1377 is an indirect
measure of the ionisation degree.  The abundance of cosmic rays is for example found to be similar to that of HCO$^+$ when CO is not depleted \citep{caselli02}.  Assuming that the vertical magnetic field is in
equipartition with the total pressure \citep[e.g.][]{ferreira11,vollmer18}, the launch conditions for an MHD wind in the inner few pc of NGC~1377 require a $B$ field of a few m$G$. This is a similar to field-strengths found in the
central region of dusty galaxies \citep{mcbride15,yoasthull19}. However,  independent measurements of the magnetic field are still missing for NGC~1377.

\subsubsection{An illustrative model}
\label{s:wind_model}

\begin{figure}[tbh]
\includegraphics[width=10.1cm, trim = 3.9cm 0cm 0cm 0cm, clip]{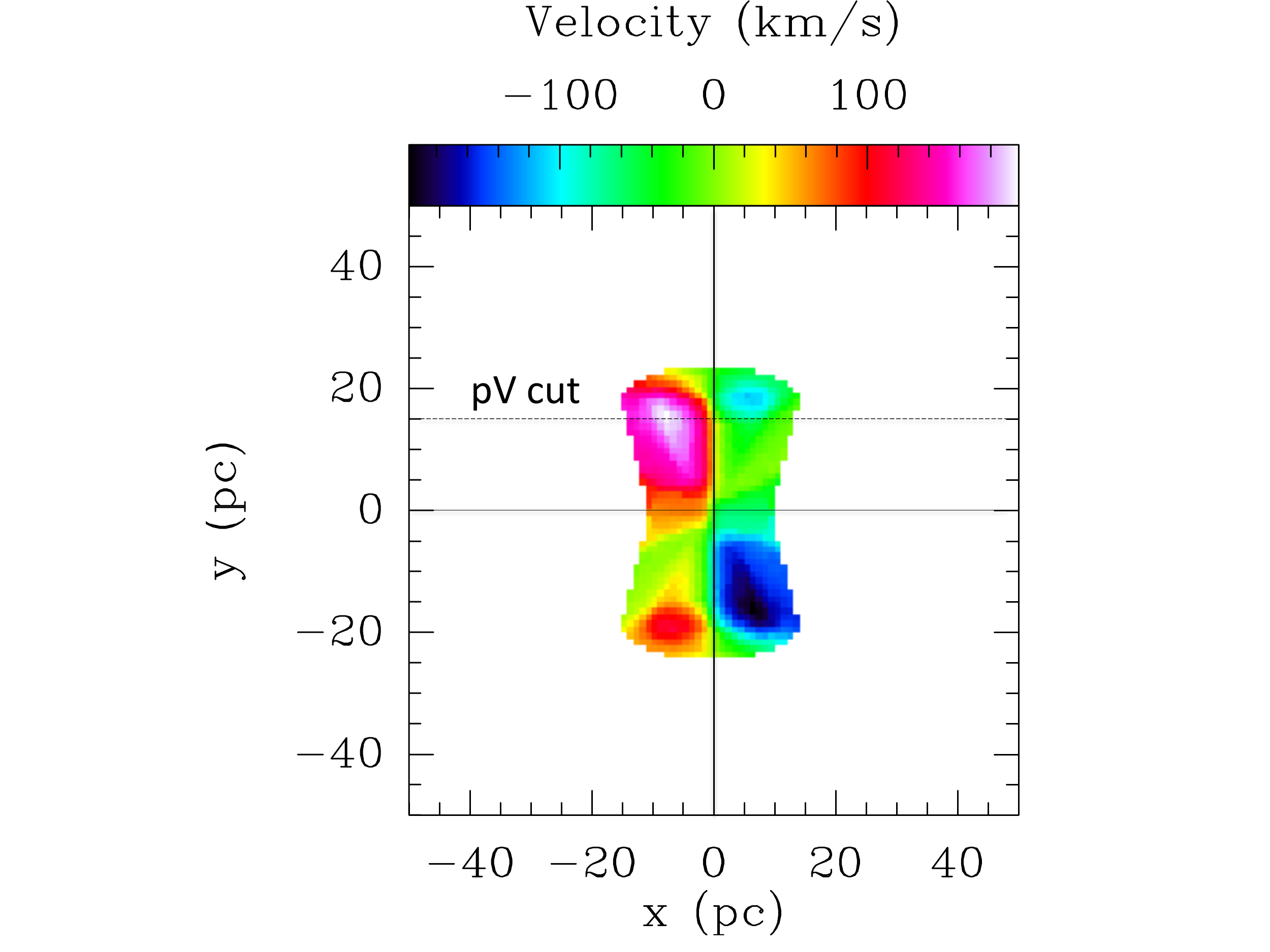}
\includegraphics[width=9.8cm, trim = 2.5cm 0cm 0cm 0cm, clip]{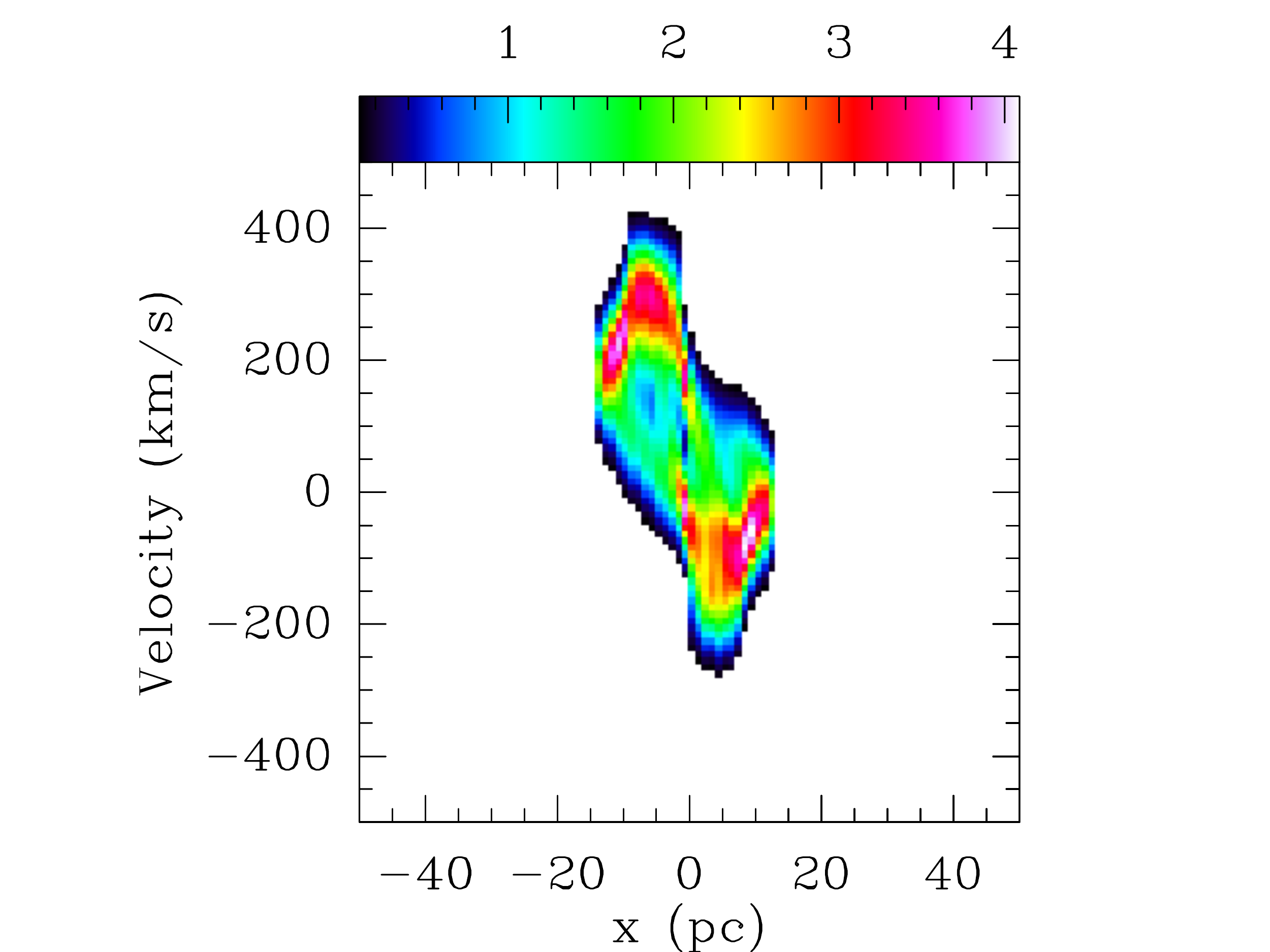}
\caption{\label{f:wind_model}  Simple, illustrative model of a rotating disk-wind. In this example, we assume a uniform nuclear disk of inclination $i$=70$^{\circ}$ in Keplerian rotation ($v_{\rm K}$) in response to an SMBH of mass
$7 \times 10^{6}$ \msun\ and a negligible disk mass. We let the outflow velocity $v_{\rm out}$=$a v_{\rm K}$ with $a$=1. We restrict the disk radius to $r$=20 pc and do not include the molecular jet in the model.
We let the wind have an inclination to the rotating disk of $\phi$=25$^{\circ}$ giving the wind an opening angle of 50$^{\circ}$. The structure is "observed" with a resolution of 1 pc. Top panel: Model moment 1 (velocity field)
map showing the resulting rotating wind. Lower panel: Model transversal pV cut at a distance 0.\asec 15 from the centre. }
\end{figure}

To illustrate how a magneto-centrifugally driven wind may appear, we have modelled a simple rotating wind emerging from a disk with outflow velocity $v_{\rm out}$ linked to 
the Keplerian rotational velocity $v_{\rm K}$ as $v_{\rm out}$=$av_{\rm K}$ (Fig.~\ref{f:wind_model}). In our model we have selected $a$ to be unity, but how the outflowing wind speed
is related to the disk rotational velocity is more complex and requires modelling to be determined. Studies indicate lower values of $a$ of 0.15-0.3 \citep[e.g.][]{ostriker97,ferreira06,kraus17}
at least as long as the distance from the disk is below the Alfv\'en surface \citet{cabrit99}. Right above the disk, the outflow
velocity is expected to be dominated by the toroidal (rotational) component, while the poloidal component is expected to dominate beyond the Alfv\'en surface. 

The illustrative example in Fig.~\ref{f:wind_model} is idealized and does not include a molecular jet component or extended gas. However, it's structure can be compared to
the inner part of the narrow wind. The wind-rotation is clear in Fig.~\ref{f:mom}, and in the panels c and d  in Fig.~\ref{f:jet_channel}. Here it is also evident that the counterpart to the
north-eastern red wind-component is missing on the blue side to the south-west. Wind-rotation signatures are lost to the south sooner than to the north.  In Fig.~\ref{f:jetcut}
we have marked the expected signatures of the rotating wind in the 0.\asec 15 transversal cut across the jet.


\subsubsection{Other, or auxiliary,  wind-driving mechanisms}

The momentum flux of the narrow wind, ${\dot M}v$$\sim$$7L/c$,  is higher than expected for a radiation pressure driven flow and the relatively low opening angle (50$^{\circ}$ - 70$^{\circ}$)  is more
consistent with an MHD wind \citep{ouyed97}.  However, we cannot exclude that radiation pressure is providing auxiliary driving. Scenarios of radiation-pressure aided  MHD jets and winds for cold disks have,
for example, been discussed by \citet{cao12} (see also \citet{vollmer18}). Since the inner region of NGC~1377 is very dusty it is primarily radiation pressure from dust that should be important.  The
misalignment between the jet and narrow wind results in jet-wind interaction (Sec.~\ref{s:bow}). Therefore, even if the wind is direct driven, the jet impacts it and is a possible source of turbulence.

The wind launched from the torus of NGC~1377 resembles to some extent the outflowing torus seen in NGC1068 \citep{garcia19} where a hot AGN wind is entraining molecular gas in the torus. 
The torus may collimate the hot wind which would then entrain cold molecular gas.  There is also the radiation-driven fountain model \citep{wada12} where an AGN-powered dusty outflow, a failed wind, and 
an inflow, form a dusty hollow cone. However,  there is  currently no evidence for the presence of a hot ionized wind, at least not emerging from the inner region of NGC 1377, nor is there any evidence of a supernova
driven wind \citep{costagliola16}. \citet{roussel06} suggest that most of the ionizing photons in the inner region of NGC~1377 are deeply buried. They do not detect any optical [O III], [O II], or  H$\beta$ emission and
infer that  [N II] and [S II] emission arise in the foreground.  \citet{roussel06} propose that a nascent ($<$1 Myr) starburst is embedded at the core of NGC~1377.  We cannot exclude that star formation is occurring inside
the opaque nuclear dust core, but there is no indication that it has enough power to drive the jet and narrow wind of NGC~1377.

 More multi-wavelength studies of the inner region of NGC~1377 are necessary to determine the true nature of the buried activity and the launch and driving mechanism of the wind. Here we suggest that the narrow wind is primarily a
magneto-centrifugal wind aided by radiation pressure, and that its structure and turbulence transmission is impacted by jet-wind interactions.

 \subsection{Driving of the slow wind}
 
As discussed in Sec.~\ref{s:bow}, the slow wind may have formed from the leading- and  internal working surfaces of the jet. The slow gas has the expected pV diagram behaviour of a jet cocoon (Sec.~\ref{s:bow}) producing
an envelope of gas. It is also possible that  that star formation outside the dusty nucleus may occur and can help drive feedback in NGC~1377.  One example of such a scenario is the enshrouded luminous infrared galaxy NGC~4418.
 An optical study suggests the presence of a kpc-scale dusty superwind, disconnected from the embedded nuclear activity \citep{ohyama19}.  A large-scale superwind in NGC~1377 is unlikely to
 drive the narrow wind or the molecular jet, but may potentially impact the slow wind.

\subsection{Jet driving mechanism}
\label{s:jet_driving}

\citet{aalto16} discussed several possible scenarios for driving the molecular jet: i) entrainment by a very faint radio jet, ii) gas accretion into the SMBH or onto nuclear disk, iii) radiation pressure from dust, or iv) a starburst wind. 
The two latter scenarios seem less likely since it is difficult to explain the extremely collimated structure in a radiation pressure powered wind and there is not enough power in star formation. 

An estimate of the momentum flux of the jet (${\dot M}v$) is $\sim$$37L/c$ (using the conservative estimate of $M$(H$_2$)) which is much higher than expected for radiation pressure driven flows, but consistent with values
found for  AGN-powered and jetted outflows \citep[e.g.][]{feruglio10,garcia14}, and also for protostellar feedback in the earlier stages of evolution \citep[e.g.][]{bontemps96}. There are several examples of molecular gas being
entrained by an AGN-driven radio jet. The gas is shocked and heated by the jet, and in some cases, also carried out of the galaxy. It is also not uncommon for the nuclear accretion disk
to be misaligned with the host galaxy and the radio jet therefore propagating through its disk  \citep[e.g.][]{matsushita07,garcia14,morganti15,dasyra15}. So far, for cases where a radio jet is entraining molecular gas, it is found 
that the gas is lining the radio jet, in contrast to the case for NGC~1377, where the cool molecular gas occupies the spine of the jet. In addition, no radio-jet has been found in NGC~1377 which is one of the most radio-quiet objects
in the sky. This suggests that the molecular material may represent the primary ejected material in NGC~1377. Such scenarios have been discussed and modelled by \citet{panoglou12}. 

If the molecular jet is not entrained by a radio jet it may in itself be an MHD disk-wind powered by accretion onto the circumnuclear disk.  The rough equipartition between gravitational and kinetic energies in the system (Sec.~\ref{s:energy})
is consistent with the notion of an outflow powered by gravitationally driven inflow on scales of 100 pc down to $<$1 pc.  The envelope (a possible jet cocoon, linked to the slow wind (Sec.~\ref{s:bow}) is a potential reservoir for a large scale
inflow, and we find evidence of an asymmetric inflow
towards the nucleus.  In addition, the jet launch region must be warped in comparison to the region from where the narrow wind is launched, and nuclear warps are associated with inflows of different angular momentum \citep[][e.g.]{ogilvie13}. The different orientation and misalignment of the jet with respect to the narrow wind appears to imply a different origin of the jet and wind. However, they may both stem from the same MHD process where a nuclear warp results in different orientations of the outflows.

Rotation of the jet/outflow is an expected feature of MHD jets and winds.  If a jet is emerging from the centre of a fast rotating disk, the outflow material is ejected with a significant
amount of angular momentum, which may become conserved along the jet streamlines. (Note, there are also other processes that can result in jet rotation \citep[e.g.][]{fendt11,soker05}).  Observational evidence for outflow rotation can be
found in protostellar jets and winds \citep[e.g.][]{matthews10,bjerkeli16,hirota17,lee17} and in winds and radio jets in AGNs  \citep[e.g.][]{mangham17,raiteri17,britzen17} - as well as in the narrow wind in NGC~1377.   The velocity in the jet
is dominated by rotation right above the disk, until the Alfv\'en surface (its location is model dependent but for some scenarios $Z_{\rm A}$$\simeq 10 \times r_0$ \citep{cabrit99}). Well above the Alfv\'en surface ($Z_{\rm A}$=1-3 pc) the jet
velocity should be mostly vertical and $v_{\rm out}$ is likely the maximum allowed velocity of $\sim$600 \kms. The jet rotation signature should be significantly lower than the jet velocity and may also be difficult to spatially resolve.  Therefore,
we adopt an estimate to the jet rotation of $\sim$60 \kms\ from the estimates of jet dispersion.

The launch region of an MHD jet can be estimated using the formalism of \citet{anderson03}.  They argue that astrophysical magnetocentrifugal jets should be capable of escaping the potential 
well of the central object fairly easily. Hence the kinetic energy of the jet is higher than the gravitational binding energy at the launch point. Inserting an outflow velocity of  $v_{\rm out}$$\sim$600 \kms,  jet radius
 $r_{\rm jet}$$\sim$3 pc, a far-field jet rotational component of 60 \kms, and the mass of the central object of $9 \times 10^6$ \msun, into their equation (5) yields a launch radius for the molecular jet of $r_{\rm launch}$$\sim$0.4 pc. 
This is inside the nuclear dust concentration,  but still more than an order of magnitude larger than the dust sublimation radius \citep{aalto17}. 

The precession time-scale of the molecular jet of NGC~1377 is relatively long, 0.42 Myr for a jet velocity of 600 \kms\ for it to be associated with the orbital time for the jet launch region implying.  \citet{aalto16} discuss possible origins of
the precession (see their Sec. 4.1.4) including a misalignment between the spin orientation of the black hole and the accretion disk, an uneven accretion flow or driving by a binary SMBH.  Further studies are required to link the cause of the 
direction changes of the jet to the jet launching mechanism. 

In some aspects, the MHD molecular jet suggested here is similar to those found in the early stages of protostellar growth. The similarity lies in the morphology and structure of the molecular jet,  as well as harbouring a compact, embedded
object. However, the nature of the central object and the time-scales involved are very different. This is expected to impact the relative properties of a protostellar and NGC~1377-like outflow in significant ways.

\subsection{Alternative jet-driving mechanisms}

The possibility of a faded or under-luminous, quenched jet entraining the gas in NGC~1377 is discussed by  \citet{aalto16}. It is possible that the SMBH of NGC~1377 is growing due to an advection dominated accretion flow
(ADAF) \citep[e.g.][]{abramowicz88,heckman14}, which would allow for high accretion rates accompanied with low radiation. ADAF flows  are, however, expected to result in powerful relativistic radio jets  \citep[e.g.][]{sadowski13} and the
extreme radio-quietness of NGC~1377 is a challenge to such a model.  Alternatively,  entrainment and acceleration may be carried out by a hot, thermal jet with a radius that is significantly smaller than that of the molecular jet, and unresolvable by our current ALMA beam. It is not clear if such a jet could harbour the necessary kinetic energy to carry out the molecular gas.


\section{The nature of the large-scale molecular disk}
\label{s:disk}

We find a $r$$\sim$60 pc molecular structure aligned with the stellar disk. However, the molecular rotation velocity is very slow, $v_{\rm rot}$$\sim$27 \kms,  much
too slow for the enclosed stellar mass ($\sim$$10^8$ \msun\ inside $r$=60 pc (estimated from HST H-band image, HST program GO14728, J Gallagher PI)) which
would require a gas disk rotation velocity of $v_{\rm rot}$=75 \kms.
A low inclination, $\sim$20$^{\circ}$, would be required for the observed velocity to be consistent with the required $v_{\rm rot}$. However,  there is no indication
of a near face-on stellar disk and according to the Hyperleda database, the inclination of NGC~1377 is $\sim$90$^{\circ}$.
The turbulent line widths ($\sigma$=30-50 \kms) in the disk are higher than the rotational velocity which suggests a thick disk that  supports itself by turbulence.
 This could also explain why no star formation has been detected in the central region of NGC~1377. The turbulence may be for example powered by returning
 gas from the wind or jet, or from a failed wind. Alternatively,  the molecular gas is not distributed in a disk, but is e.g. a filamentary structure possibly infalling onto
 the central region of NGC~1377.


\section{Potential  link between feedback and growth in NGC~1377}
\label{s:growth}

The wind and jet of NGC~1377 may constitute a different type of outflow than what is usually found towards AGNs (or starbursts). There is, as of yet, no evidence of
a significant hot wind or a radio jet in NGC~1377.  Instead the gas may be driven out by a molecular, magneto-centrifugal rotating wind and jet. In this scenario, the outflows are not
a direct result of the radiative or mechanical feedback of an AGN, but instead arise as an effect of inflowing gas and a magnetic field. These winds and jets help accretion through
removing angular momentum. The estimated obscuration is relatively high ($N$(H$_2)$$ \simeq$$ 1.8 \times 10^{24}$$\cmmt$)
on scales $r$= 1.5 pc and the suggested presence of an inflow is consistent with a nucleus still in unstable growth.  If the narrow wind+jet molecular mass of  $\sim1 \times 10^7$ \msun\ recently resided in the inner
$r$$\sim$3 pc the obscuration would have been up to two orders of magnitude higher - rivaling the extreme obscuration of the Compact Obscured Nuclei (CONs) \citep{costagliola13,sakamoto13,aalto15b,aalto19}.
It is possible that NGC~1377 is caught in a post-CON evolutionary phase, where a recent growth spurt with associated wind-outflow activity has removed a significant fraction of obscuring material. 
However, it is also possible that the suggested link between inflow and outflow in NGC~1377 prevents CON-like obscuring layers to form.

The estimated SMBH mass, $9^{+2}_{-3} \times 10^6$ \msun\ inside $r$=1.4 pc, indicates that it is massive with respect to the stellar velocity dispersion.  NGC~1377 is on the upper bound of the $M-\sigma$ relation
\citep[e.g.][]{graham11,greene16} that links stellar velocity dispersion to black hole mass. 

The jet and wind of NGC~1377 are also different through being roughly perpendicular to the stellar disk and not at an oblique angle, unlike known radio-jet driven outflows in, for example, NGC~1068, IC5063,
ESO 420-G13 or NGC613 \citep[e.g.][]{garcia14, dasyra15, audibert19, fernandez20}. The molecular jet of NGC~1377 may have more in common with the collimated molecular outflows recently discovered towards galaxies
with CONs \citep[e.g.][]{falstad18,barcos18,falstad19}. The jet and narrow wind are surrounded by a slow wind of gas which may partially be stemming from internal and external shocks in the jet producing a
cocoon or an envelope of gas (Sec.~\ref{s:bow}). This material has low angular momentum and low-velocity which means that it is readily available to fuel nuclear growth.

We find tentative evidence of inflowing gas (on several scales) that could help power the disk-wind and fuel the nuclear growth. Further observations and modelling
are required to determine the inflow rate and the origin of the inflowing gas. It is possible that it is coming from the enveloping cocoon with an origin in the wind and jet gas. if so,
the SMBH of NGC~1377 has a substantial supply of gas that, eventually if star formation is prevented, can serve as fuel for SMBH growth. Such a feeding-cycle could also provide part of the answer to the question of
the origin of the molecular gas concentration \citep{aalto12b}. In this ageing, post-starburst galaxy, a circulation system may explain why there is still a significant reservoir of molecular gas.

The nuclear luminosity implies a relatively high current accretion state of NGC~1377, but it is not enough to significantly grow the SMBH.   With the estimated, radiative,
accretion rate, the SMBH would grow with only $10^{-3} - 10^{-2}$ \msun\  per year and, it would take at least 100 Myr for the SMBH to double its mass. However, it is likely that the nuclear luminosity and accretion rate was
significantly higher in the recent past, giving rise to the molecular jet and wind. Alternatively,  NGC~1377 is, or recently has been, in a state of advective dominated accretion where the energy is not radiated.


\section{Conclusions}

With high-resolution ($0.\asec 02 \times 0.\asec 03$ ($2 \times 3$ pc)) ALMA 345~GHz observations of CO 3--2,  HCO$^+$ 4--3, vibrationally excited HCN 4--3 $\nu_2$=1f, and
continuum we have studied the remarkable, radio-quiet, and collimated molecular jet and wind of the lenticular galaxy NGC~1377. 

The morphology and structure of the jet and wind is resolved, revealing clumpy,  turbulent emission with high gas dispersion ($\sigma$$>$40 \kms) and a radial excitation
gradient in the gas.  The jet is narrow with an average width of 3-7 pc, but it undergoes an apparent expansion 20-40 pc from the nucleus where the jet broadens to 10-17 pc. This widening of the jet may be caused by bow shocks
from jet-wind interactions or internal working surfaces stemming from variations in the outflow speed. The molecular jet has a length of 150 pc and appears to be launched from $r\lapprox$0.4~pc.  It shows velocity
reversals that we propose are either due to regular precession or more episodic directional changes of ejection.   We estimate the precession angle to 15$^{\circ}$ and the jet outflow speed to $\sim$600 \kms. 

The jet has a high momentum flux ${\dot M}v$=${\dot p}$ of $\sim$37$L/c$ which is in general consistent both with AGN feedback and momentum fluxes found for early stage protostellar feedback.
However, the luminosity of the nuclear source may have varied considerably during the duration of the outflow, rendering the value of ${\dot p}$ uncertain.
We propose that the molecular jet is a gravitationally powered magneto-centrifugal wind, but cannot exclude entrainment by a hitherto undetected, weak radio or thermal jet. There
is enough kinetic energy in the jet to power the high levels of turbulence in the wind and disk. Jet-wind interactions in the form of bow-shocks are possible processes of transfer. 

The jet is surrounded by a narrow molecular wind of opening angle 50$^{\circ}$ - 70$^{\circ}$, which is is launched from the inner ($r$=1.5-3 pc) region of NGC~1377. We find strong evidence of wind-rotation and suggest
that also the narrow wind is a magneto-centrifugal wind, and that it appears to be primarily direct driven. Its outflow velocities are  $90 \pm 30$  \kms\ with
a significant momentum flux of ${\dot p}$=7$L/c$. Radiation pressure cannot be excluded as an auxiliary driving force,  but the relatively narrow opening angle is consistent with a magnetically confined
wind. The narrow wind and jet are misaligned which is expected from the signatures of nuclear warping and the misalignment can result
in jet-wind interactions. 

The jet and narrow wind are enveloped by a slow wind which we propose  is a cocoon,  forming from the jet-wind interaction or from gas ejected in internal working surfaces of the jet.

An $r$=2 pc, asymmetric, torus-like dust structure is found in 0.8~mm continuum. It also shows compact emission from vibrationally excited HCN. The nuclear dust emission is hot with
inferred $T_{\rm d}$$>$180 K and is near Compton Thick with an estimated gas column density of $N$(H$_2)$$\simeq$$1.8 \times 10^{24}$$\cmmt$. We estimate the luminosity of the nuclear dust source
to $\sim$$4.8 \times 10^9$ \lsun. Absence of signs of ongoing star formation lead us to suggest that it stems from accretion onto the dust-enshrouded SMBH. The nuclear high-velocity gas is consistent with 
dynamical mass of $9^{+2}_{-3} \times 10^6$ \msun\ inside $r$=1.4 pc.  The lopsided, obscuring torus appears to be a dynamically active region, harbouring both in- and outflows.
High-velocity blue-shifted gas (in CO and H$^{13}$CN) located 2 pc west of the nucleus may be a signature of gas on highly non-circular orbits, but the presence of a
second nucleus can not yet be dismissed. Larger scale ($r$=60 pc) emission aligned with the stellar disk is rotating too slowly for its enclosed mass and may be a pressure-supported, turbulent disk (or is not
actually a molecular disk.) 

We suggest that the SMBH of NGC~1377 is in a state of moderate growth, at the end of a more intense phase of activity and possibly also from a state of more extreme nuclear obscuration. The SMBH of NGC~1377
appears massive with respect to its stellar velocity dispersion and this opens new questions on SMBH growth processes in obscured, dusty galaxies. The nuclear growth may be fueled by low-angular momentum gas inflowing from the gas
ejected in the molecular jet and wind. In this scenario, the molecular jet, wind and potential inflow represent a feedback loop of cyclic, central accretion which may be an important  phase in the evolution of NGC~1377. In this
ageing, post-starburst galaxy, such a circulation system may explain why there is still a significant reservoir of molecular gas. Further studies are required to investigate if an over-massive SMBH is an expected consequence of an
outflow-powered feedback loop.

\begin{acknowledgements}
This paper makes use of the following ALMA data:
    ADS/JAO.ALMA\#2012.1.00900.S. ALMA is a partnership of ESO (representing
    its Member States), NSF (USA), and NINS (Japan), together with NRC
    (Canada) and NSC and ASIAA (Taiwan), in cooperation with the Republic of
    Chile. The Joint ALMA Observatory is operated by ESO, AUI/NRAO, and
NAOJ.
We thank the Nordic ALMA ARC node for excellent support. The Nordic ARC node is
funded through Swedish Research Council grant No 2017-00648. S.A. gratefully acknowledges
support from an ERC Advanced Grant 789410 and from the Swedish Research Council.
K.S. was supported by grant MOST 102-2119-M-001-011-MY3
S.G.B. acknowledges the economic support from grants PGC2018-094671-B-I00
and AYA2016-76682-C3-2-P (MCIU/AEI/FEDER,UE). A.S.E. was supported by NSF grant
AST 1816838. J.S.G. thanks the University of Wisconsin-Madison Foundation and College of
Letters \& Science for partial support of this work.  K.M.D. acknowledge financial support by the
Hellenic Foundation for Research and Innovation (HFRI), project number 188.
This research has made use of the NASA/IPAC Extragalactic Database (NED) which is operated by the
Jet Propulsion Laboratory, California Institute of Technology, under
contract with the National Aeronautics and Space Administration.

\end{acknowledgements}

\bibliographystyle{aa}
\bibliography{n1377_ALMA_b7HR}

\end{document}